\documentclass[twocolumn]{jpsj4}

\usepackage{amsmath}
\usepackage{amsfonts}
\usepackage{amssymb}
\usepackage{txfonts}
\usepackage{bm}
\usepackage{tabularx}
\usepackage[dvipdfm]{graphicx}
\usepackage{arydshln}
\usepackage[dvipdfm]{color}

\usepackage{times}

\def\Vec#1{\bm{#1}}

\newcommand{\sub}[1]{$_{\mathrm {#1}}$}
\newcommand{\subm}[1]{_{\mathrm {#1}}}

\newcommand{\SRO}{Sr\sub{3}{Ru\sub{2}O\sub{7}}}

\renewcommand{\deg}{^{\circ}}
\newcommand{\Tc}{T\subm{c}}

\newcommand{\Ic}{I\subm{c}}

\newcommand{\Hcc}{H\subm{c2}}

\newcommand{\sro}{Sr\sub{2}{RuO\sub{4}}}

\newcommand{\bra}[1]{\lvert#1\rangle}

\newcommand{\uu}{\bra{\uparrow\uparrow}}
\newcommand{\ud}{\bra{\uparrow\downarrow}}
\newcommand{\du}{\bra{\downarrow\uparrow}}
\newcommand{\dd}{\bra{\downarrow\downarrow}}
\newcommand{\dv}{\bm{d}}
\newcommand{\xh}{\hat{x}}
\newcommand{\yh}{\hat{y}}
\newcommand{\zh}{\hat{z}}
\newcommand{\tu}{\theta_\uparrow}
\newcommand{\td}{\theta_\downarrow}
\newcommand{\esp}{\bra{\Psi}_\mathrm{ESP}}

\newcommand{\eq}[1]{eq.~\eqref{#1}}

\title{Evaluation of Spin-Triplet Superconductivity in \sro}

\author{
	Yoshiteru \textsc{Maeno}$^{1}$, 
	Shunichiro \textsc{Kittaka}$^{1,2}$, 
	Takuji \textsc{Nomura}$^{3}$, 
	Shingo \textsc{Yonezawa}$^{1}$, and
	Kenji \textsc{Ishida}$^{1}$
}

\inst{$^{1}$Department of Physics, Graduate School of Science,
		Kyoto University, Kyoto 606-8502, Japan\\
	$^{2}$Institute for Solid State Physics, University of Tokyo, Kashiwa, Chiba 277-8581, Japan\\
	$^{3}$Condensed Matter Science Division, Japan Atomic Energy Agency, Sayo, Hyogo 679-5148, Japan\\
}

\recdate{\today}

\abst{
This review presents a summary and evaluations of the superconducting properties of the layered ruthenate \sro\ as they are known in the autumn of 2011. 
This paper appends the main progress that has been made since the preceding review by Mackenzie and Maeno was published in 2003. 
Here, special focus is placed on the critical evaluation of the spin-triplet, odd-parity pairing scenario applied to \sro. 
After an introduction to superconductors with possible odd-parity pairing, accumulated evidence for the pairing symmetry of \sro\ is examined. 
Then, significant recent progress on the theoretical approaches to the superconducting pairing by Coulomb repulsion is reviewed. 
A section is devoted to some experimental properties of \sro\ that seem to defy simple explanations 
in terms of currently available spin-triplet scenario. 
The next section deals with some new developments using eutectic boundaries and micro-crystals, 
which reveals novel superconducting phenomena related to chiral edge states, odd-frequency pairing states, and half-fluxoid states. 
Some of these properties are intimately connected with the properties as a topological superconductor. 
The article concludes with a summary of knowledge emerged from the study of \sro\ 
that are now more widely applied to understand the physics of other unconventional superconductors, 
as well as with a brief discussion of relatively unexplored but promising areas of ongoing and future studies of \sro. 
}

\kword{Sr$_2$RuO$_4$, ruthenate, spin-triplet superconductivity, topological superconductor}

\begin{document}

\maketitle

\makeatletter
    \renewcommand{\theequation}{%
    \arabic{equation}}
    \renewcommand{\thefigure}{%
    \arabic{figure}}
    \renewcommand{\thetable}{%
    \Roman{table}}
\makeatother
\section{Spin-triplet superconductors}
\subsection{Candidates of Spin-triplet superconductors}

In the last three decades, and particularly since the discovery of 
high-transition-temperature (high-$\Tc$) superconductivity of the cuprates \cite{Bednorz1986ZPB}, 
studies of ``unconventional'' superconductivity have been one of the main topics in condensed-matter physics. 
Here we designate the term ``unconventional'' as the pairing based on non-phonon mechanisms. \cite{comment1}
The unconventional superconductivity is mainly found in heavy-fermion superconductors (since 1978),\cite{Steglich1979PRL} 
high-$\Tc$ cuprates (since 1986),\cite{Bednorz1986ZPB} organic superconductors (since 1980),\cite{Jerome1980JPL} ruthenate superconductors (since 1994),\cite{Maeno1994Nature}
and iron-pnictide superconductors (since 2008).\cite{Kamihara2009JACS}
Unconventional superconductivity is characterized by the anisotropic gap function or order parameter 
which is integrated to be zero or a small value due to the variations of the wave function ``phase'',
in contrast to an ordinary $s$-wave state. 
In many of them, including high-$\Tc$ cuprates and iron pnictides, 
the electrons are clearly paired in \textit{spin-singlet} states. 
In this point of view, they are similar to conventional $s$-wave superconductors, 
in which the spin-degrees of freedom is lost in the charged superfluids. 
\textit{Spin-triplet} superfluid states are fully established in the Fermi liquid $^3$He \cite{Wheatley1975RMP,Leggett1975RMP}, 
for which spin and mass supercurrents emerge in the charge-neutral superfluids. 
The question is then whether or not spin-triplet \textit{superconductors} exist, 
and what novel superconducting properties they may exhibit due to their charge and spin supercurrents.  

There are several classes of candidates of spin-triplet superconductors represented in Table~\ref{tripletSC}. 
We should note here that due to strong spin-orbit interactions of heavy elements, 
spin is not a good quantum number in the $f$-electron based heavy fermions; 
their possible pseudo-spin-triplet pairing may be better termed as ``odd-parity'' pairing for their orbital wave function symmetry. 
Keeping this in mind, we will nevertheless use the term ``spin-triplet'' whenever it causes no confusion. 
Among the heavy fermions, UPt$_3$ is the leading candidate of a triplet superconductor \cite{Stewart1984PRL}. 
The measurements of spin susceptibility indicate that pairing in UNi$_2$Al$_3$ \cite{Geibel1991ZPB} is also triplet \cite{Ishida2002PRL}, 
whereas the pairing in UPd$_2$Al$_3$ is clearly spin-singlet \cite{Kyogaku1993JPSJ,Tou2005JPSJ}. 
Recent progress in the studies of ferromagnetic heavy-fermion superconductors 
such as UGe$_2$ \cite{Saxena2000Nature}, URhGe \cite{Aoki2001Nature}, and UIr \cite{Akazawa2004JPSJ}, and in particular UCoGe \cite{Huy2007PRL}, 
motivates researchers to seek for
obtaining direct evidence for spin-triplet pairing.

\begin{table*}[tb]
\begin{center}
\caption{Selection of candidates of spin-triplet superconductors. HF: heavy fermion superconductors, 
NCS: Noncentrosymmetric superconductors, FM: ferromagnetic superconductors, $\ast$: superconductivity under pressure.}
{\renewcommand\arraystretch{1}
\begin{tabular}{cccc}
\hline \hline
Materials & Classification & Spin evidence of triplet pairing & Properties \\ \hline
$^3$He & Superfluid & magnetization, NMR etc.\cite{Wheatley1975RMP} & $p$-wave, A phase is chiral \\ \hdashline
\sro & Oxide & NMR, polarized neutron & 2D analogue of $^3$He-A \\
 & & & Chiral $p$-wave \\ \hdashline
UPt$_3$ & HF & NMR\cite{Tou1998PRL} & $f$-wave \\
UBe$_{13}$, URu$_2$Si$_2$, UNi$_2$Al$_3$ & HF & NMR\cite{Tou2005JPSJ} & \\
UGe$_2^\ast$, URhGe, UCoGe & FM, HF & Indirect & Anomalous $\Hcc$\cite{Sheikin2001PRB,Hardy2005PRL,Levy2005Science,Aoki2009JPSJ} \\
UIr$^\ast$ & NCS,FM,HF & Indirect & \\
CeIrSi$_3^\ast$ & NCS, HF & NMR\cite{Mukuda2010PRL} &  \\
Li$_2$Pt$_3$B & NCS & NMR\cite{Nishiyama2007PRL} &  \\
CePt$_3$Si & NCS, HF & Indirect &  \\
CeRhSi$_3^\ast$ & NCS, HF & Indirect &   Anomalous $\Hcc$\cite{Kimura2007PRL} \\
S/FM/S & junctions & Indirect ($\Ic$)\cite{Keizer2006Nature,Anwar2010PRB,Robinson2010Science,Khaire2010PRL} & Odd-freq., even-parity, $s$-wave \\ \hline \hline
\label{tripletSC}
\end{tabular}}
\end{center}
\end{table*}

Among the organic superconductors, possible spin-triplet pairing in quasi-one-dimensional superconductors 
(TMTSF)$_2X$ ($X$ = ClO$_4$ and PF$_6$) has been considered. \cite{Lebed1986JETPL,Lee1975RMP}
However, spin-singlet pairing is demonstrated by nuclear magnetic resonance (NMR) Knight shift for $X$ = ClO$_4$ \cite{Shinagawa2007PRL}. 
In addition, recent results of detailed field-orientation dependence of the onset temperature derived from resistivity indicate that 
the unusual survival of superconductivity in high magnetic fields is attributable to the formation of the Fulde-Ferrell-Larkin-Ovchinnikov (FFLO) state of spin-singlet superconductivity \cite{Yonezawa2008PRL,Yonezawa2008JPSJ}. 

A new class of spin-triplet pairing has been recognized 
in some superconductors having crystal structures without inversion symmetry, such as Li$_2$Pt$_3$B \cite{Nishiyama2007PRL}. 
It is believed that if the spin-orbit splitting of the Fermi surfaces 
due to internal electric fields exceeds the superconducting gap energy, 
the pairing state is no longer classified as purely spin-singlet or triplet, but their mixing occurs \cite{Frigeri2004PRL,Fujimoto2007JPSJ}. 
As a result, a variety of unusual behavior characteristic of spin-triplet pairing such as extremely high upper critical field $\Hcc$ emerges. 
Examples are CePt$_3$Si,\cite{Bauer2004PRL} CeRhSi$_3$,\cite{Kimura2005PRL} etc. 
In particular, anisotropic spin susceptibility in the superconducting state has been reported for CeIrSi$_3$.\cite{Mukuda2010PRL} 
As Mukuda \textit{et al}. explained, 
if the spin susceptibility is dominated by strong Rashba-type spin-orbit coupling, 
Knight-shift measurements do not give direct evidence for the admixture between the Cooper-pairing states of even and odd parity.
The anisotropy may well be dominated by that of Van-Vleck-like susceptibility.

\sro, the main subject of this review, is the first oxide superconductor in the same layered structure as high-$\Tc$ cuprates 
but with a low superconducting transition temperature of 1.5~K \cite{Maeno1994Nature,Mackenzie2003RMP}. 
The availability of high-quality single crystals and the relative simplicity of its fully-characterized Fermi surfaces 
promoted a large number of experimental as well as theoretical studies. 
\sro\ is now established as one of the archetypal unconventional superconductors. 
In particular, it is considered most probably as a spin-triplet superconductor, 
comparable with the odd-parity, pseudo-spin-triplet superconductor UPt$_3$ \cite{Leggett2006QL}.

We should add that the emergence of spin-triplet pairing states is also discussed 
in superconductor-ferromagnet-superconductor (S/FM/S) junctions. 
By the proximity effect into ferromagnetic half-metal such as CrO$_2$, 
$s$-wave pair amplitude may penetrate as spin-triplet $s$-wave state \cite{Keizer2006Nature,Anwar2010PRB}.
Such an unusual state is known as odd-frequency pairing \cite{Tanaka2007PRL}. 
Recently, the long-range proximity effect due to odd-frequency spin-triplet currents is also demonstrated 
using ferromagnetic hybrid layers with some disorder and rotation of magnetization. \cite{Robinson2010Science, Khaire2010PRL}

\subsection{Unconventional superconducting properties of \sro}
From its Fermi surface topography and $\Hcc$ anisotropy, 
\sro\ is characterized as a highly anisotropic superconductor with the anisotropy ratio of 20. 
Nevertheless, the out-of-plane coherence length $\xi_c$ is several times greater than 
the interlayer distance to allow interlayer coherence. 
Table~\ref{SC_para} summarizes basic superconducting parameters obtained mainly from the anisotropic $\Hcc$ and the specific heat. 
We note that the evaluation of $\xi_c$ assumes orbital limiting of $\Hcc$. 

\sro\ exhibits a number of phenomena reflecting its unconventional superconductivity.
The extreme sensitivity to non-magnetic impurities and defects indicates non-$s$-wave pairing with sign-changes on the superconducting gap.
In fact the transition temperature $\Tc$ diminishes at the impurity concentration 
at which the mean impurity-impurity distance becomes comparable to the in-plane coherence length $\xi_{ab}$, 
consistent with the modified Abrikosov-Gorkov prediction \cite{Mackenzie1998PRL,Mao1999PRB}.
Absence of coherence peak (Hebel-Schlichter peak) and the power-law temperature dependence in the NMR spin-lattice relaxation rate $1/T_1$ 
indicate non-$s$-wave with highly anisotropic gap structure. \cite{Ishida2008JPCS}
The magnetic field dependence of the specific heat at low temperatures reveals a plateau 
in the intermediate $H$, indicative of the multi-gap superconductivity reflecting the existence of three Fermi surfaces with different orbital characters. \cite{Deguchi2004PRL}

\begin{table}
\begin{center}
\caption{Superconducting parameters for \sro. 
$H_\mathrm{c}$ is the thermodynamical critical field,
$\xi$ is the coherence length, $\lambda$ is the penetration depth, and
$\kappa$ is the Ginzburg-Landau parameter. 
$^{\ast}$Conventional orbital depairing for $\Hcc$ is assumed.
}
\renewcommand{\arraystretch}{1}
\begin{tabular}{cccccc}
\hline \hline
Parameter & & & $ab$ & $c$ \\ \hline
$\Tc$ & (K) & 1.50 & & \\
$\mu_0\Hcc(0)$ & (T) & & 1.50 & 0.075 \\
$\mu_0H_\mathrm{c}(0)$ & (T) & 0.023 & & \\
$\xi(0)$ & ($\AA$) & & 660 & $33^{\ast}$ \\
$\lambda(0)$ & ($\AA$) & & 1900 & $3.0 \times 10^4$ \\
$\kappa(0)$ & & & 2.6 & 46 \\
$\xi_{ab}/\xi_c$ & & 20 & & \\ \hline
\label{SC_para}
\end{tabular}
\end{center}
\end{table}

\begin{table*}[tb]
\begin{center}
\caption{Selection of candidates of spin-triplet superconductors. HF: heavy fermion superconductors, 
NCS: Noncentrosymmetric superconductors, FM: ferromagnetic superconductors, $\ast$: superconductivity under pressure.}
{\renewcommand\arraystretch{1}
\begin{tabular}{cccc}
\hline \hline
Materials & Classification & Spin evidence of triplet pairing & Properties \\ \hline
$^3$He & Superfluid & magnetization, NMR etc.\cite{Wheatley1975RMP} & $p$-wave, A phase is chiral \\ \hdashline
\sro & Oxide & NMR, polarized neutron & 2D analogue of $^3$He-A \\
 & & & Chiral $p$-wave \\ \hdashline
UPt$_3$ & HF & NMR\cite{Tou1998PRL} & $f$-wave \\
UBe$_{13}$, URu$_2$Si$_2$, UNi$_2$Al$_3$ & HF & NMR\cite{Tou2005JPSJ} & \\
UGe$_2^\ast$, URhGe, UCoGe & FM, HF & Indirect & Anomalous $\Hcc$\cite{Sheikin2001PRB,Hardy2005PRL,Levy2005Science,Aoki2009JPSJ} \\
UIr$^\ast$ & NCS,FM,HF & Indirect & \\
CeIrSi$_3^\ast$ & NCS, HF & NMR\cite{Mukuda2010PRL} &  \\
Li$_2$Pt$_3$B & NCS & NMR\cite{Nishiyama2007PRL} &  \\
CePt$_3$Si & NCS, HF & Indirect &  \\
CeRhSi$_3^\ast$ & NCS, HF & Indirect &   Anomalous $\Hcc$\cite{Kimura2007PRL} \\
S/FM/S & junctions & Indirect ($\Ic$)\cite{Keizer2006Nature,Anwar2010PRB,Robinson2010Science,Khaire2010PRL} & Odd-freq., even-parity, $s$-wave \\ \hline \hline
\label{tripletSC}
\end{tabular}}
\end{center}
\end{table*}

It should be mentioned that the theoretical proposal for the similarity 
between the superconductivity of \sro\ and the spin-triplet superfluidity of $^3$He 
soon after the discovery of ruthenate superconductivity \cite{Rice1995} strongly motivated 
the experimental efforts to determine the symmetry of unconventional superconductivity in \sro.
This leads to the first direct experimental evidence of spin-triplet pairing in \sro\ 
by the measurements of the electron spin susceptibility with NMR \cite{Ishida1998Nature}. 
Combined with the observation of internal magnetic field by $\muup$SR, \cite{Luke1998}
this result yields a complete description of the superconducting state of \sro\ as 
a two-dimensional analogue of the $^3$He-A phase, i.e., a spin-triplet chiral $p$-wave state. \cite{Rice1998}

\subsection{Scope of this review}
There have been several review articles devoted to the superconductivity of \sro \cite{Mackenzie2003RMP,Sigrist2005PTPS,Kallin2009JPCM, Ishida2008JPCS}. 
In addition, we mention a review on the Fermi-liquid properties in the normal state \cite{Bergemann2003AP} and 
stories behind the discovery of its superconductivity.\cite{Lichtenberg2002PSSC, Maeno2011}
Especially, an extensive review of the developments up to the spring of 2002 by Mackenzie and Maeno has been cited more than 500 times.
Since then, much research progress has been made on \sro\ in both experiments and theories, 
especially concerning the superconducting gap structure, the order-parameter symmetry, 
and the properties of superconducting states in junctions and at boundaries. 

The purpose of this review is to make a critical examination of the spin-triplet pairing scenario of \sro, 
in order to fully understand the behavior of such superconducting state. 
The ideas emerged from the studies of \sro\ include multi-gap superconductivity of multi-band systems, 
physics of superconductivity with time-reversal symmetry breaking, pairing mechanism in highly correlated materials
in which spin fluctuation is not necessarily strong. 
Such analysis should help understanding the properties of other unconventional superconductors, 
and moreover help resolving some of the long standing remaining issues of spin-triplet superfluid $^3$He \cite{Leggett2006QL}.

In \S~2, we first introduce the order parameter of spin-triplet superconductivity, the $d$-vector, 
especially the $d$-vector states relevant to the superconductivity of \sro.
We examine the experimental evidence for spin-triplet pairing based on traditional probes, but with innovative approaches. 
A variety of NMR Knight-shift experiments and polarized neutron experiments establish the spin-triplet pairing. 
We also address alternative possibilities of the $d$-vector orientation consistent with the NMR. 
We then describe experiments establishing the time-reversal symmetry breaking, or chiral nature, of the order parameter. 

Section~3 summarizes the theoretical advancement in unconventional pairing mechanism by Coulomb repulsion. 
The importance of non-boson exchange, namely vertex-correction processes, is emphasized. 
We also review and examine other approaches to the unconventional pairing mechanism for \sro. 
The important roles played by the spin-orbit interaction, which determines the direction of the $d$-vector, are also discussed. 

Section~4 deals with the superconducting properties of \sro\ which remain unresolved within the current theoretical models. 
These are the origin of the strong suppression of $\Hcc$ with the fields parallel to the RuO$_2$ plane and 
the emergence of a second superconducting phase at low temperatures under the same condition. 

Section~5 describes novel superconducting phenomena in \sro, such as enhanced superconductivity (the so-called 3-K phase) 
at the interface between \sro\ and Ru in the eutectic system, and novel proximity effects in various superconducting junctions involving \sro. 
Descriptions of Andreev bound states in terms of odd-frequency pairing and chiral edge states are introduced. 
Many of these phenomena are characteristics of a topological superconductor.
We also describe the current experimental status of observing half-fluxoid states in micron-scale rings of \sro\ crystals and 
discuss the physics of half-quantum vortices in spin-triplet superconductors. 

Finally in \S~6, we summarize the present status towards establishing the spin-triplet superconductivity in \sro\ and 
discuss the future prospects of the physics of topological superconductivity.

\section{Evidence for spin triplet pairing}
\subsection{Order parameter of spin-triplet superconductivity: the $d$-vector}

We start with a useful description of the order parameter of spin-triplet superconductivity, 
namely the $d$-vector. 
The superconducting order parameter is given by a function of spin and momentum 
in general: $\Delta_{\sigma\sigma'}(\mib{k})$. 
Roughly speaking, $\Delta_{\sigma\sigma'}(\mib{k})$ is regarded as a wave function 
of a Cooper pair formed by two quasi-particles 
whose momenta and spins are ($\mib{k}$, $\sigma$) and ($-\mib{k}$, $\sigma'$). 
The magnitude of the total spin of a Cooper pair is $S=1$ in spin-triplet pairing, 
in contrast to $S=0$ in spin-singlet pairing. 
In spin-triplet pairing, the degrees of freedom about the direction of the spin $S=1$ 
remain even in the superconducting state. 
For spin-triplet pairing, $\Delta_{\sigma\sigma'}(\mib{k})$ is expressed 
using the $d$-vector as \cite{Balian1963}: 
\begin{equation}
\Delta_{\sigma\sigma^\prime}(\mib{k}) ={\rm i} [ ( \mib{d}(\mib{k}) \cdot \mib{\sigma} ) \sigma_y ]_{\sigma\sigma'}, 
\end{equation}
or
\begin{align}
&\left(
\begin{array}{cc}
\Delta_{\uparrow\uparrow}(\mib{k}) & \Delta_{\uparrow\downarrow}(\mib{k})\\
\Delta_{\downarrow\uparrow}(\mib{k}) & \Delta_{\downarrow\downarrow}(\mib{k})
\end{array} \right) \nonumber \\ &= 
\left(
\begin{array}{cc}
-d_x(\mib{k})+{\rm i} d_y({\mib{k}}) & d_z(\mib{k})\\
d_z(\mib{k}) & d_x(\mib{k})+{\rm i} d_y({\mib{k}})
\end{array} \right), 
\end{align}
where $\mib{\sigma}$ is the Pauli spin matrix vector: 
$\mib{\sigma}=(\sigma_x, \sigma_y, \sigma_z)$. 
The order parameter of triplet pairing states is described completely 
by the three components of the $d$-vector $\mib{d}(\mib{k})$. 
$\mib{d}(\mib{k})$ is an odd-parity function of $\mib{k}$. 
The $d$-vector behaves as a vector under rotations in spin space. 
Here we should note that the direction of the $d$-vector is perpendicular 
to the total spin $S$ of a Cooper pair. 
To see this simply, we consider the case of $\mib{d} \parallel z$, i.e., $\mib{d}(\mib{k})=(0,0,d_z(\mib{k}))$. 
Then we have $\Delta_{\uparrow\uparrow}(\mib{k})=\Delta_{\downarrow\downarrow}(\mib{k})=0$, 
i.e., $S_z=0$, which means $\mib{d} \perp \mib{S}$. 
In general, each component of the $d$-vector ($d_x$, $d_y$, or $d_z$) represents the pair amplitude with the Cooper-pair spin
perpendicular to the corresponding basis ($\hat{x}$, $\hat{y}$, or $\hat{z}$).
Triplet pairing states satisfying $\mib{d}(\mib{k}) \times \mib{d}^*(\mib{k}) = 0$ 
($\mib{d}(\mib{k}) \times \mib{d}^*(\mib{k}) \neq 0$) are called {\it unitary} 
({\it non-unitary}) states. 
In non-unitary states, spins of the Cooper pairs are polarized. 
We can see this simply by considering the quantity 
$|\Delta_{\uparrow\uparrow}(\mib{k})|^2-|\Delta_{\downarrow\downarrow}(\mib{k})|^2 
= 2 {\rm i} [\mib{d}(\mib{k}) \times \mib{d}^*(\mib{k})]_z \neq 0$ for non-unitary states.

The order parameters of superfluid phases of $^3$He are expressed as:\cite{Vollhardt}
$\mib{d}(\mib{k}) \propto \hat{z}(k_x \pm {\rm i}k_y)$ (A phase), 
$\mib{d}(\mib{k}) \propto \hat{x}k_x+\hat{y}k_y+\hat{z}k_z$ (B phase), 
$\mib{d}(\mib{k}) \propto (\hat{x}+{\rm i}\hat{y})(k_x + {\rm i}k_y)$ (A1 phase emerging under magnetic fields).
Both of the A and B phases are unitary states, 
while the A1 phase is a non-unitary state with spin polarization. 
For the state $\mib{d}(\mib{k}) \propto \hat{z}(k_x \pm {\rm i} k_y)$, 
we can choose a new spin coordinate system 
in which $\Delta_{\uparrow\downarrow}(\mib{k}) = \Delta_{\downarrow\uparrow}(\mib{k})=0$, i.e., 
$d_z({\mib k})=0$ at all ${\mib k}$. 
To see this, we rotate the spin coordinate system 
by $2\pi/3$ around the $[111]$ axis. 
Then the basis vectors $\hat{x}$, $\hat{y}$ and $\hat{z}$ are 
transfered to $\hat{z}$, $\hat{x}$ and $\hat{y}$, respectively, and consequently 
we have $d_z({\mib k})=d_x({\mib k})=0$ for all ${\mib k}$ 
in the new spin coordinate system. 
In general, if we can choose appropriate spin coordinates 
by which $d_z({\mib k})=0$ for all ${\mib k}$, then 
we call the pairing state ``equal-spin pairing'' (ESP) state. 
The A and A1 phases are ESP states, while the B phase is not.
For those not familiar with the description of the spin-triplet order parameter, 
the review by Mackenzie and Maeno (\S~IV and Appendix~D), as well as the Appendix of this review, gives 
an introductory but more detailed treatment in terms of the $d$-vector. 
For example, one can find the relation between the spin bases and the $d$-vector bases, and
the expression of the superconducting gap in terms of the $d$-vector.

\begin{table*}[tb]
\caption{
List of possible five $d$-vector states~\cite{Rice1995} for \sro.  
Irreducible representations of the $C_{4v}$ symmetry group are displayed 
by both of the Bethe and Mulliken notations in the first two columns. 
The minus sign in the Bethe notation and the index ``$u$'' (denoting {\it ungerade})
in the Mulliken notation represent odd parity. 
TRSB in the fifth column means time-reversal symmetry breaking. 
$\Gamma_1^-$ and $\Gamma_5^-$ are analogous states to the $^3$He-B 
and $^3$He-A phases, respectively. 
$\Gamma_5^-$ corresponds to the chiral $p$-wave state.
In the absence of the spin-orbit interaction among Ru-$4d$ electrons, 
i.e., assuming isotropy in spin space, 
any superpositions of these five states are also possible. 
Note that the column ``$d$-vector'' displays only the symmetry 
of $\mib{d}(\mib{k})$ as a function of $\mib{k}$. 
For example, if we take the functions $\sin k_x$ and $\sin k_y$ for the $k_x$- 
and $k_y$-symmetry, then we have $\mib{d}(\mib{k}) \propto \hat{z}(\sin k_x \pm {\rm i} \sin k_y)$ 
for $\Gamma_5^-$. All of the five states are unitary ESP states, 
and give full energy gap at the Fermi level. }
\label{table3}
\begin{center}
\begin{tabular}{ccccccc}
Bethe & Mulliken & $d$-vector & direction of $\mib{d}$ & TRSB & Analogy to $^3$He \\ \hline
$\Gamma_1^-$ & $A_{1u}$ & $\hat{x} k_x +  \hat{y} k_y $ & $ \mib{d} \parallel ab$ & No & BW state (B phase) \\
$\Gamma_2^-$ & $A_{2u}$ & $\hat{x} k_y -  \hat{y} k_x $ & $ \mib{d} \parallel ab$ & No & \\
$\Gamma_3^-$ & $B_{1u}$ & $\hat{x} k_x -  \hat{y} k_y $ & $ \mib{d} \parallel ab$ & No & \\
$\Gamma_4^-$ & $B_{2u}$ & $\hat{x} k_y +  \hat{y} k_x $ & $ \mib{d} \parallel ab$ & No & \\
$\Gamma_5^-$ & $E_u$    & $\hat{z} (k_x \pm {\rm i} k_y)$ & $\mib{d} \parallel c$ & Yes & ABM state (A phase) 
\end{tabular}
\end{center}
\end{table*}

\begin{figure}
\begin{center}
\includegraphics[width=3in]{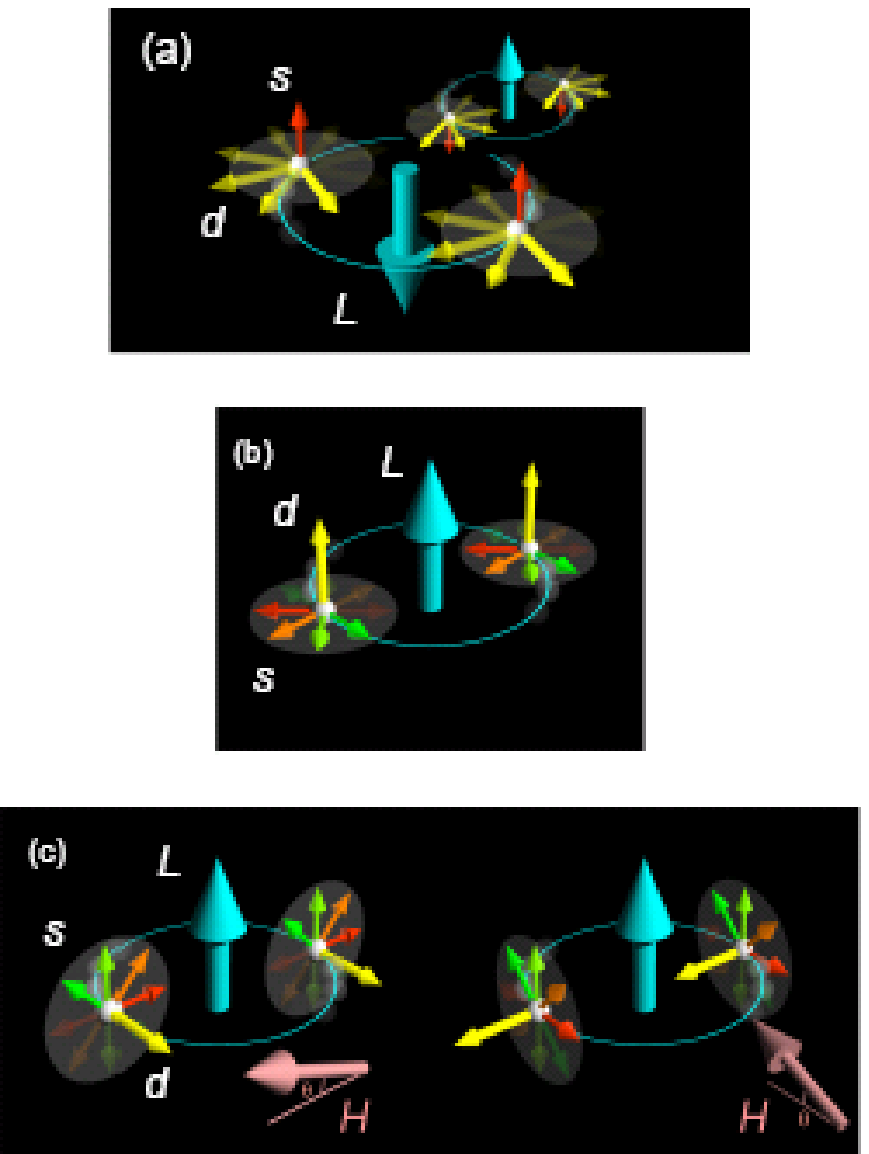}
\caption{
(Color online) 
Illustrations of Cooper-pair spins and orbital angular momenta for spin-triplet superconductivity. 
The Cooper-pair spins $S$ are depicted by small arrows in various colors; 
the $d$-vector shown by yellow arrows is defined perpendicular to the spins. 
The orbital angular momentum $L$ of a Cooper pair is shown by a large blue arrow. 
(a) The $S$ and $L$ vectors for the $\Gamma_1$ state, $\Vec{d}(\mib{k}) = \Delta_0(\hat{x}k_x + \hat{y}k_y)$. 
It can be thought of as a quantum-mechanical superposition of $S_z=+1$ and $S_z=-1$ states (see text).
(b) The promising ground state $\Vec{d}(\mib{k})=\hat{z}\Delta_0(k_x + {\rm i} k_y)$ with the $d$-vector pointing in the $c$ direction. 
(c) Possible states under magnetic fields. 
The $d$-vector rotates within the basal plane in response to the external magnetic fields. 
The Cooper-pair spins can be polarized by magnetic fields of any directions. 
}
\label{2-1_d-vector}
\end{center}
\end{figure}

Down to the low temperatures,
\sro\ maintains a tetragonal structure with the crystal point group symmetry $D_\mathrm{4h}$. \cite{Chmaissem1998PRB}
Neglecting the dispersion along the out-of-plane $c$ direction, 
possible spin-triplet states are limited to those for the two-dimensional square lattice with $C_\mathrm{4v}$ symmetry and are given in Table~\ref{table3}. 
Figures~\ref{2-1_d-vector}(a) and \ref{2-1_d-vector}(b) depict the $\Gamma_1$ and $\Gamma_5$ states, respectively. 
The former can be thought of as 
the quantum-mechanical superposition of the two states with antiparallel combinations of spin and orbital wave functions,
with quenched overall orbital angular momentum: 
\begin{equation}
\Vec{d}(\mib{k}) \propto (\hat{x}+{\rm i}\hat{y})(k_x-{\rm i}k_y)+(\hat{x}-{\rm i}\hat{y})(k_x+{\rm i}k_y)
\end{equation}
The $d$-vector $\hat{x} \pm {\rm i}\hat{y}$ lies in the basal plane, 
which for illustrative purpose is depicted by yellow arrows in Fig.~\ref{2-1_d-vector}(a). 
The $\Gamma_5^-$ states in Table~\ref{table3} are called the {\it chiral pairing} states, 
because these states possess two polarizations of relative orbital angular momentum of pairing quasi-particles: 
left- and right-handed polarizations correspond to $k_x + {\rm i} k_y$ and $k_x - {\rm i} k_y$, respectively. 
They are the states with the orbital angular momentum $L_z=+1$ or $-1$ as represented by big arrows in Fig.~\ref{2-1_d-vector}(b); 
the Cooper-pair spins lie in the basal plane as $\Vec{d} \perp \Vec{S}$. 
We note that all five states $\Gamma_1$ - $\Gamma_5$ are the so-called ESP states,
for which the spin state of the Cooper pair is always a linear superposition of $S_z=+1$ and $S_z=-1$ states 
by a suitable choice of spin axes.

\subsection{Spin part of the order parameter}

The direct evidence for spin-triplet pairing in \sro\ is based mainly on the following two kinds of experiments. 
One is the electronic spin susceptibility measurements by the NMR Knight shifts of both $^{17}$O and $^{99}$Ru nuclei \cite{Ishida1998Nature,Mukuda1998JPSJ}.
The other evidence comes from the measurements of local magnetization by polarized neutron diffraction; 
the so-called Shull-Wedgewood experiments \cite{Duffy2000PRL}. 

In this subsection, we focus on the NMR experiments in some detail.
The NMR Knight shift measures the effective field at the nucleus produced by the electrons, 
and thus is related to the microscopic spin susceptibility at the NMR nuclear site. 
Since the spin susceptibility in the superconducting state cannot be extracted from the bulk susceptibility due to the Meissner shielding, 
the Knight-shift measurement is considered as the most reliable method 
to measure the spin susceptibility in the superconducting state \cite{Ishida2008JPCS}. 

\begin{figure}
\begin{center}
\includegraphics[width=2.7in]{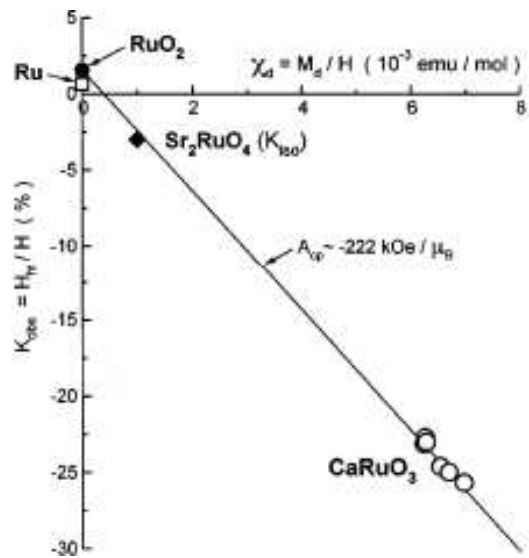}
\caption{
Knight shift vs bulk susceptibility determined in the range $T = 1.4$ - 20~K for a weakly-correlated metal RuO$_2$, 
an electron-correlated superconductor \sro, and a correlated, nearly ferromagnetic metal CaRuO$_3$.
The hyperfine coupling constant due to the inner core polarization $A_\mathrm{cp} \sim - 22.2$~T/$\mu_\mathrm{B}$ 
is estimated from a linear slope with the assumption that $K_\mathrm{orb} =1.59$\% is common among these ruthenates. 
The value of $A_\mathrm{cp}$ is close to $A_\mathrm{hf} \sim 30$~T/$\mu_\mathrm{B}$ in the FM state of SrRuO$_3$. 
A large negative $K$ value of \sro\ indicates that its susceptibility is dominated 
by the $d$-electron spin susceptibility (Fig.~5 in Ref.~\ref{Mukuda1999PRB}). 
}
\label{2-1_K-chi}
\end{center}
\end{figure}

\begin{figure}
\begin{center}
\includegraphics[width=2.5in]{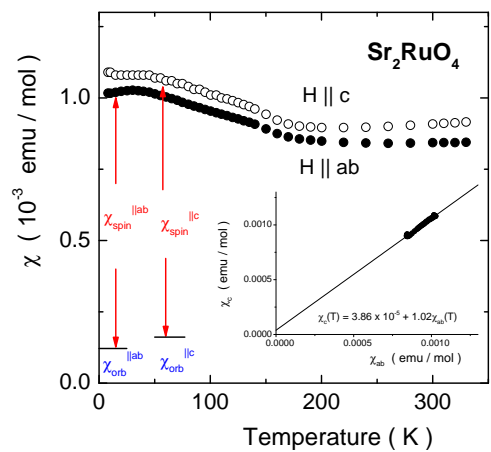}
\caption{(Color online) 
Bulk susceptibility of \sro\ (Refs. \ref{Ishida2001} and \ref{Murakawa2007JPSJ}) illustrating the spin and orbital contributions (see text for details).
}
\label{2-1_susceptibility}
\end{center}
\end{figure}

The Cooper pair spin susceptibility of \sro\ can be reliably extracted from the NMR Knight shift $K$ for the following reasons. 
First, the NMR and nuclear quadrupole resonance (NQR) measurements have been performed on a variety of nuclei, 
three different crystal sites of $^{17}$O (NMR and NQR), $^{99}$Ru (NMR), and $^{101}$Ru (NQR), 
covering wide parameter range in the field-temperature ($H$--$T$) phase diagram. 
Thus accidental signal cancellation due to the structure factor is avoided. 
In addition, it is noticeable that one can also obtain the bulk susceptibility and nuclear spin-lattice relaxation rate ($1/T_1$) within the same measurement setup. 
This fact enables us to carefully check the consistency of the data from different viewpoints.
Second, overheating of the sample by RF pulses is avoided by monitoring the intensity of the NMR peak itself, 
thereby using the NMR nuclei as an internal thermometer (Supplement of Ref.~\ref{Ishida1998Nature}). 
Third, based on the systematic studies of $K$ among different ruthenates, i.e., RuO$_2$, \sro, and CaRuO$_3$, 
it is clear that a large negative value of $K$ for \sro\ originates from 
the contribution from core polarization due to $d$-electrons (Fig.~\ref{2-1_K-chi}).\cite{Mukuda1999PRB} 
From such analysis, the contribution from the orbital part is estimated as 1.0\% 
while the spin contribution is $-4.5$\% for $K \parallel c$ of \sro. 
With the anisotropy of the spin part of only 2\% as extracted from the $\chi_{ab}$ vs $\chi_c$ plot, \cite{Ishida1997PRB}
the dominance of the spin part in the susceptibility \cite{Ishida2001,Murakawa2007} is illustrated in Fig.~\ref{2-1_susceptibility}.
Such properties of \sro\ ensure that its NMR Knight shift serves as a reliable high-sensitivity probe for the Cooper pair spin susceptibility. 

Ishida and coworkers have measured Knight shifts $K(T)$ at the Ru and O sites in high-quality single crystals. 
Measurements were performed using a $^3$He-$^4$He dilution refrigerator with the sample crystals 
directly immersed in liquid helium to ensure good thermal contact to the bath. 
As for the $^{17}$O Knight-shift measurements, they assigned NMR signals arising from the different crystal sites (planer and apical O sites), 
and estimated the spin part of the Knight shift above $\Tc$ at each site. 
They investigated temperature dependence of the Knight shift at the planer O site, 
where the spin density is 6 times larger than that in the apical O site, 
and showed that its Knight shift remains unchanged in the field range 
($0.35~{\rm T} < \mu_0H < 1.1$~T) applied parallel to the RuO$_2$ plane. \cite{Ishida1998Nature}

\begin{figure}
\begin{center}
\includegraphics[width=2.5in]{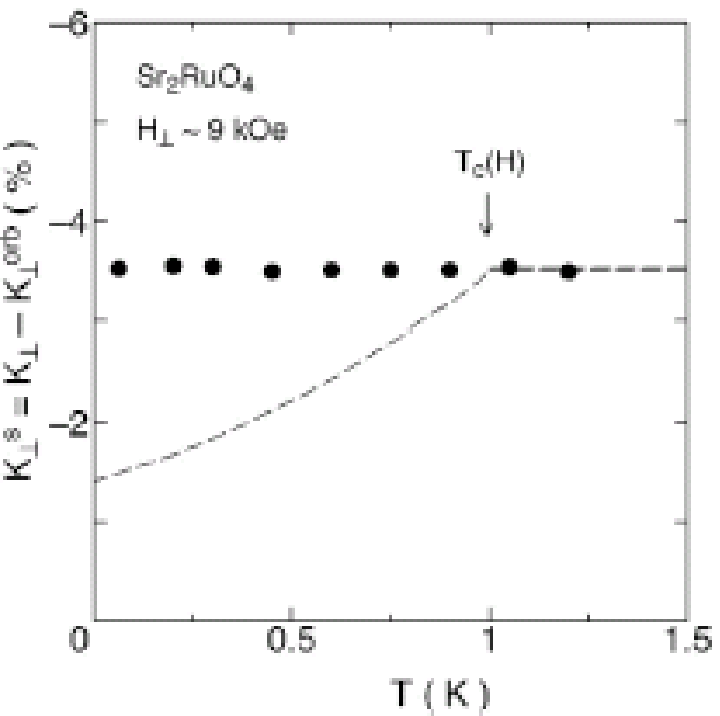}
\caption{
Knight shift of $^{99}$Ru nuclei of \sro\ with the applied field parallel to the basal $ab$ plane (Ref.~\ref{Ishida2001}). 
The spin part $K_s(T)$, after subtracting the temperature-independent orbital part $K_\mathrm{orb}$ from the total shift $K(T)$, is plotted.
}
\label{2-1_Knight}
\end{center}
\end{figure}

They also measured the Knight shift at the Ru site in the RuO$_2$ plane. 
The advantage of the Ru Knight shift measurements is that the value of the $^{99}$Ru Knight shift 
is an order of magnitude larger than that of the $^{17}$O Knight shift, 
since the $^{99}$Ru shift is strongly affected by Ru-$4d$ electronic spins 
through the large hyperfine coupling constant of $A_\mathrm{cp} = -25$~T/$\mu_\mathrm{B}$ originating from the core polarization effect, 
where $\mu_\mathrm{B}$ is the Bohr magneton. 
This means that the Ru Knight shift can detect much smaller changes of the Ru-$4d$ electronic state 
when superconductivity sets in. The temperature dependence of the $^{99}$Ru shift was measured 
in the fields of 0.68, 0.9, and 1.05~T parallel to the RuO$_2$ plane and the spin part of the Knight shift ($^{99}K_\mathrm{spin} = K(T) -K_\mathrm{orb}$) 
obtained at 0.9~T is shown in Fig.~\ref{2-1_Knight}. 

$^{99}K_\mathrm{spin}$ does not change on passing through $\Tc \sim 1.0$~K at $\mu_0H \sim 0.9$~T. 
If a spin-singlet $d$-wave state with a line-node gap were realized, 
the $T$ dependence of $^{99}K_\mathrm{spin}$ would behave as drawn by the dashed curve in Fig.~\ref{2-1_Knight}. 
Here we adopted the magnitude of the gap $2\Delta_0=4k_\mathrm{B}\Tc$, which has been used for analyses of various physical quantities.
It is obvious that the unchanged Knight shift in the superconducting state at both the O and Ru sites in the RuO$_2$ plane 
cannot be understood by the spin-singlet scenarios, 
but strongly suggests the Cooper-pair spins are in the triplet state and can be polarized to the applied field parallel to the RuO$_2$ plane.

In general, the spin susceptibility in the spin-triplet superconducting state is expected to show anisotropic temperature dependence 
reflecting the magnitude and direction of the applied field. 
When the interaction acting on the triplet pairs is strong enough to lock the spins of a superconducting pair in a certain direction of the crystal, 
the anisotropy of the spin susceptibility, i.e., the anisotropic behavior of the Knight shift should be observed in the superconducting state. 
The Knight shift is unchanged when magnetic fields are parallel to the spin direction ($\Vec{d} \perp H$). 
But when magnetic fields are applied perpendicular to the spin direction ($\Vec{d} \parallel H$), 
the Knight shift decreases depending on the superconducting gap structure. 
On the other hand, if the energy of the pinning interaction is weaker so that the spins can reorient easily toward the applied magnetic field, 
the Knight shift is invariant in any field direction. 
Thus the Knight shift measurements in small fields are important to understand the pinning interaction of the $d$-vector.

\begin{figure}
\begin{center}
\includegraphics[width=3in]{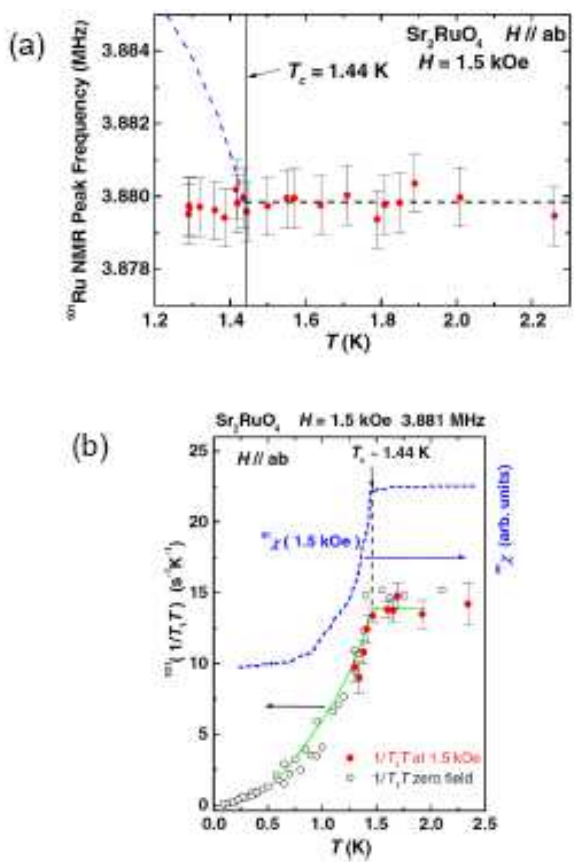}
\caption{(Color online) 
(a) Low-field Knight shift of $^{101}$Ru nuclei of \sro\ with the field parallel to the basal $ab$ plane. 
The Knight shift was obtained from the spectral shift of the NQR frequencies in the presence of a magnetic field. 
(b) Ac susceptibility $\chi_\mathrm{ac}$ and nuclear spin-lattice relaxation rate $1/T_1$ obtained in the identical setup as used for Fig.~\ref{2-1_Knight2}(a). 
Taken from Ref.~\ref{Murakawa2007JPSJ}.
}
\label{2-1_Knight2}
\end{center}
\end{figure}

In order to measure the Knight shift in such small fields, Murakawa \textit{et al}. employed a $^{101}$Ru NQR spectrum 
for observing the signals, and measured the Knight shift at the Ru site \cite{Murakawa2004PRL,Murakawa2007} as shown in Fig.~\ref{2-1_Knight2}(a). 
They measured the temperature dependence of the resonance frequency of the Ru NQR signal affected by the applied field parallel to the RuO$_2$ plane. 
The expected decrease of the spin susceptibility in the case of the spin-singlet pairing or spin-triplet 
with the $d$-vector in the RuO$_2$ plane is shown by the dotted curve. Obviously such suppression was not observed, 
and the spin susceptibility is invariant through $\Tc$ instead. 
The same measurement was done in field as small as 55~mT confirming the invariance of the spin susceptibility. 
The occurrence of the superconductivity in the field was confirmed by the measurements of the Meissner signal 
as well as $1/T_1$ of Ru using the identical set up of the Knight-shift measurements, as shown in Fig.~\ref{2-1_Knight2}(b). 
Note that the Knight-shift spectra were obtained within 100~$\muup$s after the RF pulse with the total duration of 100~$\muup$s. 
In contrast, $1/T_1$ was obtained within 100~ms after the pulse. 
Absence of excessive heating by the RF pulses for the NMR measurements was carefully confirmed by 
examining the temperature dependence of the spectral peak intensities. 

The Knight-shift measurements in magnetic fields perpendicular to the RuO$_2$ plane (parallel to the $c$ axis), denoted as $K_c$, 
were performed in the same way. However the measurements in this direction is difficult for the following two reasons: 
(1) $\Hcc$ along the $c$ axis is only 75~mT, 20 times smaller than $\Hcc$ along the RuO$_2$ plane; 
(2) the superconducting fraction with penetrating magnetic field is much more suppressed 
since the Ginzburg-Landau parameter is as small as 2.6 for $H \parallel c$. 
Thus, the measurements for $H \parallel c$ were performed with utmost care.

\begin{figure}
\begin{center}
\includegraphics[width=3.2in]{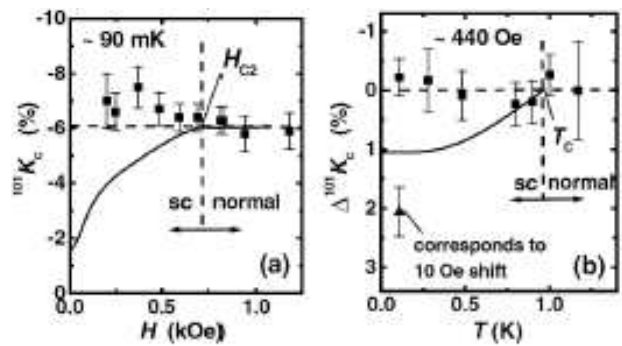}
\caption{
$^{101}$Ru Knight shift $^{101}K_c$ of \sro\ (a) at 90~mT for $H \parallel c$ as small as 20~mT and 
(b) at various temperatures at 44~mT (Ref.~\ref{Murakawa2004}). 
One can see in (a) that the orbital contribution to $K_c$ is -1.6\%. 
In (b) the data point in a triangle is obtained by intentionally varying the external field by 1~mT. 
}
\label{2-1_Knight3}
\end{center}
\end{figure}

Figure~\ref{2-1_Knight3}(a) shows $K_c$ measured at $T=90$ mK in the field range from 20~mT to 120~mT, and Fig.~\ref{2-1_Knight3}(b) 
shows the temperature dependence of $K_c$ measured at 44~mT. 
In both measurements, no appreciable changes can be detected beyond the experimental error. 
The solid curves in Figs.~\ref{2-1_Knight3}(a) and \ref{2-1_Knight3}(b) are conceivable dependences, 
when the superconducting $d$-vector is locked along the $c$ axis. 
Such dependences are inconsistent with experimental results: 
The spin susceptibility is unchanged also for magnetic fields parallel to the $c$ axis. 

\begin{figure}
\begin{center}
\includegraphics[width=3.2in]{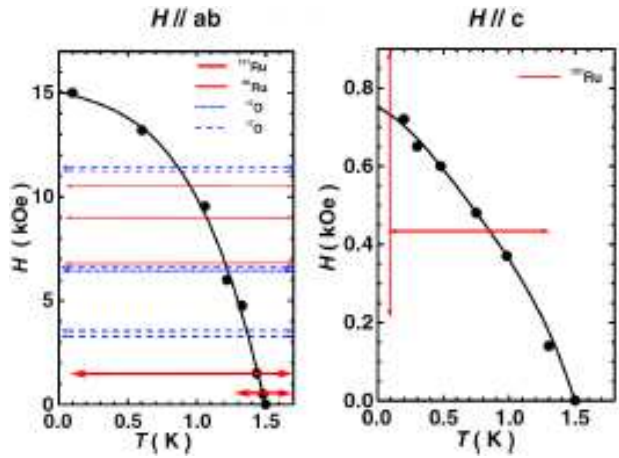}
\caption{(Color online) 
Field-temperature phase diagram of \sro\ summarizing the temperature or field sweeps of the Knight-shift measurements reported so far 
(Ref.~\ref{Murakawa2007JPSJ}).
}
\label{2-1_Knight4}
\end{center}
\end{figure}

Figure~\ref{2-1_Knight4} summarizes the temperature and field ranges of the Knight-shift measurements performed so far; 
no appreciable suppression has been detected in any of the measurements. 
One of the plausible interpretations is that the $d$-vector orients perpendicular to the external fields greater than 20~mT, 
irrespective of the field direction. 

\subsection{Chiral order parameter} \label{Sec_chiral}

\begin{figure}[h]
\begin{center}
\includegraphics[width=3.4in]{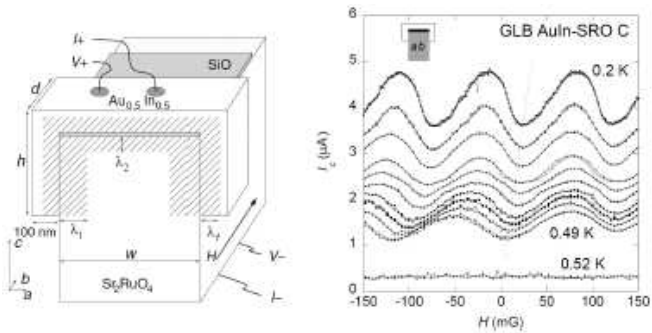}
\caption{
A $\pi$-junction SQUID consisting of Au$_{0.5}$In$_{0.5}$ and Sr$_2$RuO$_4$ \cite{Nelson2004Science}. 
The minimum of its critical current extrapolates to zero towards the superconducting $T_{\rm c}$ of an $s$-wave superconductor Au$_{0.5}$In$_{0.5}$, 
where the induced flux due to the asymmetry of the SQUID is minimized.
}
\label{FigA}
\end{center}
\end{figure}

The odd parity of the orbital part of the order parameter has been unambiguously demonstrated 
by phase-sensitive measurements by Nelson \textit{et al}. \cite{Nelson2004Science,Rice2004Science} 
The critical current of Au$_{0.5}$In$_{0.5}$-Sr$_2$RuO$_4$ superconducting quantum interference devices (SQUIDs) 
containing a loop of conventional $s$-wave superconductor Au$_{0.5}$In$_{0.5}$ and Sr$_2$RuO$_4$ prepared on Sr$_2$RuO$_4$ single crystals 
was found to be a maximum for devices with junctions on the same side of the crystal and a minimum for devices with junctions on opposite sides, 
in the limit of zero magnetic flux as shown in Fig.~\ref{FigA}. 
These findings indicate that the phase of the superconducting order parameter in Sr$_2$RuO$_4$ changes by $\pi$ under inversion. 
This result verifies the odd-parity pairing symmetry corresponding to the formation of spin-triplet Cooper pairs in Sr$_2$RuO$_4$.

Among the candidate odd-parity spin-triplet pairing states listed in Table~\ref{table3},
the $\Gamma_5$ state with the orbital order parameter $\Delta(\mib{k})=\Delta_0(k_x \pm {\rm i}k_y)$
is the so-called ``chiral superconducting state'',
which is accompanied by the broken time-reversal symmetry (TRS).
The term ``chiral'' is used for the TRS broken state associated with the orbital part of the wave function. 
The TRS breaking was first demonstrated by the emergence of internal magnetic fields 
observed in the muon spin resonance experiments \cite{Luke1998}. 
Since the specific heat demonstrates no residual density of states (DOS), the broken TRS state associated with the spin part, 
the so-called ``non-unitary state'', is unlikely realized in \sro\ 
at least under zero magnetic field.

\begin{figure}
\begin{center}
\includegraphics[width=2.8in]{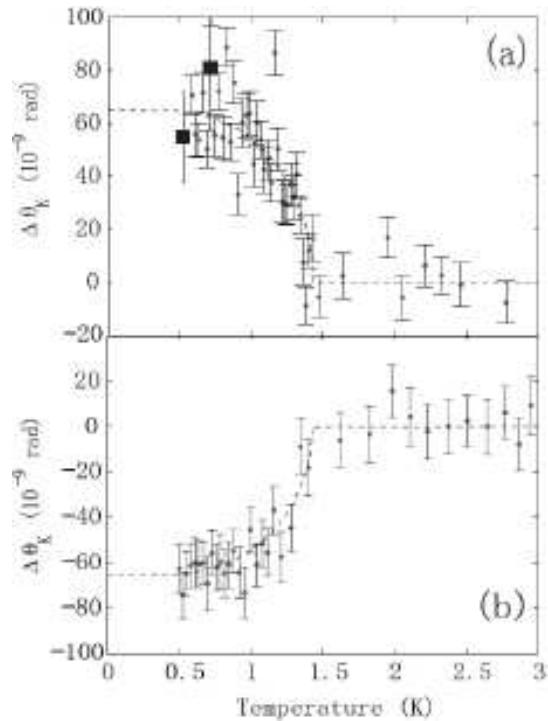}
\caption{
Rotation angle $\Delta\theta_\mathrm{K}$ of the reflected light from the surface of a \sro\ crystal 
due to magneto-optic Kerr effect (MOKE). 
The change of about 60 nrad below $\Tc$ is attributed to the broken time-reversal symmetry of chiral superconductivity. 
The sign of $\Delta\theta_\mathrm{K}$ can be manipulated by the direction of the external field applied before the measurements, 
interpreted as the alignment of chiral superconducting domains. From Xia \textit{et al}. (2008).\cite{Xia2006PRL}
}
\label{2-2_Kerr}
\end{center}
\end{figure}

An important progress has been made by the development of ultra-high sensitive magneto-optic Kerr effect (MOKE) apparatus 
based on a Sagnac interferometer \cite{Xia2006APL}. 
The rotation of the polarization plane of the reflected light below $\Tc$ shown in Fig.~\ref{2-2_Kerr} 
indicates the broken TRS in the superconducting \sro \cite{Xia2006PRL}. 
It is shown that the Kerr signal does not couple strongly with the vortices and 
that the sign of the Kerr signal is reversed by the application of the field over about 5~mT 
before the measurements under zero-field are performed. 
The latter is attributed to the need for training the chiral domains, 
i.e., enlarging the size of the chiral domain having the same sign of the orbital angular momentum of the Cooper pairs as the external field. 
Considering the size of the focused laser light used, 
the trained chiral domain size is estimated to be on the order of a few micrometers. 
It is worth mentioning that the new technique of MOKE experiments has successfully been applied to investigate broken TRS 
associated with the inverse proximity effect in the bilayer films consisting of an $s$-wave superconductor and a ferromagnet \cite{Xia2009PRL}.

The MOKE measurements for Sr$_2$RuO$_4$ stimulated 
a number of theoretical investigations. \cite{Yakovenko2007PRL,Goryo2008PRB,Roy2008PRB,Lutchyn2009PRB}
In general, the Kerr rotation angle is expressed in terms of the 
ac Hall conductivity $\sigma_{xy}(\omega)$, i.e., 
the off-diagonal part of the ac conductivity tensor. 
Using appropriate parameters for Sr$_2$RuO$_4$, 
Yakovenko evaluated the Kerr rotation angle $\Delta \theta_K$ 
for the chiral $p$-wave pairing state, and 
obtained $\Delta \theta_K \approx 230$ nrad~\cite{Yakovenko2007PRL}. 
Goryo presented a mechanism of impurity-induced Kerr rotation. 
He calculated the Hall conductivity within the low-order expansion 
in the impurity strength, and found that the leading 
contribution is given by similar diagrams to the skew-scattering 
diagrams in the extrinsic anomalous Hall effect. 
In this impurity-induced mechanism, the MOKE would be suppressed 
or would be zero for any unitary pairing states other than the chiral pairing state. 
The rotation angle $\Delta \theta_K$ is evaluated to be about 30 nrad 
for a realistic impurity mean distance $\ell=5000$~$\AA$, in agreement 
with the experimental values~\cite{Goryo2008PRB}. 
These studies on MOKE suggest that the chiral pairing state is realized 
in Sr$_2$RuO$_4$. 

The chiral superconducting state of Sr$_2$RuO$_4$ is supported by a number of experiments, 
including $\mu$SR, Kerr effect, and Josephson effect. 
The chirality, or the TRS breaking in the orbital wave function, 
is believed to originate from the $k_x \pm {\rm i}k_y$ wave function associated with the orbital angular momentum $L_z=1$ of Cooper pairs. 
In the bulk of Sr$_2$RuO$_4$ crystals, it is expected that domain structures of superconducting regions with opposite chirality form. 
Strong evidence for the presence of such order parameter domains as well as 
for the domain wall motion were obtained through anomalous behavior of diffraction patterns of Josephson junctions 
on single faces of the crystals by Kidwingira \textit{et al}. \cite{Kidwingira2006Science} 
The observed telegraph noise in the critical current as a function of magnetic field or time, 
responsible also for hysteresis observed in field sweeps of the critical current, 
is attributable to transitions between the chiral states of a domain or the motion of domain walls separating them. 
Examining the modulation envelopes the data, they estimated average domain width of $\sim 1$ $\muup$m. 
The presence of such domains confirms the broken time-reversal symmetry nature of the superconducting pairing state in Sr$_2$RuO$_4$. 

At a boundary of a chiral superconducting mono-domain, 
net superconducting edge current is expected within a healing length proportional to the superconducting coherence length. 
However, this edge current is expected to be screened by the Meissner current extending to the penetration depth from the edge. 
Since the Ginzburg-Landau parameter is $\kappa=2.6$, this cancellation must be rather significant. 
Nevertheless, within the parameters of Sr$_2$RuO$_4$, 
it is expected that the net magnetic field due to chiral edge current is as large as 1~mT. 
On the cleaved surface of a single crystal of Sr$_2$RuO$_4$, 
it is expected that the chiral edge current exists at the bulk edge 
as well as along the boundaries of $k_x+{\rm i}k_y$ ($L_z=+1$) and $k_x-{\rm i}k_y$ ($L_z=- 1$) domains. 

Aiming at the direct observation of the magnetic field due to the chiral edge current, 
measurements of scanning SQUID microscope were performed \cite{Kirtley2007PRB,Hicks2010PRB}. 
In both experiments, spontaneous magnetic field was not observed at the bulk boundaries or 
as an internal pattern suggesting the presence of chiral domain boundaries. 
After various domain configurations are considered, 
it is concluded that 
for the domain size of 2 microns or greater, the magnitude of the magnetic field has to be less than 1\% of that expected.
Another approach in search for the chiral edge current was taken by a torque magnetometry on a micron-size \sro\ ring; 
but the result indicated no edge current of the expected magnitude \cite{Jang2011Science}. 

We note that Dolocan \textit{et al}. also made direct imaging of magnetic flux structures using a scanning micro-SQUID force microscope, 
anticipating preferential penetration as well as pinning of vortices along the domain walls \cite{Dolocan2005PRB}. 
They observed that as field parallel to the c-axis is increased, individual vortices coalesce and form flux domains. 
These observations imply the existence of a mechanism for bringing vortices together overcoming the conventional repulsive vortex-vortex interaction. 
Simulation study of the magnetization process was performed by Ichioka \textit{et. al.} for the multidomain state in a chiral $p$-wave superconductor, 
using the time-dependent Ginzburg-Landau theory \cite{Ichioka2005PRB}. 
They derived a vortex sheet structure forming along a domain wall, and at higher fields the motion of the domain walls 
so that the unstable domains shrink to vanish. 
Therefore, the single domain structure would be realized at higher fields.

To explain the absence of magnetic field associated with the edge current, some scenarios have been proposed. 
Raghu \textit{et al}. proposed an alternative triplet pairing state, \cite{Raghu2010}
focusing on the hidden quasi-one-dimensional character of the $xz$ and $yz$ bands 
and the interplay between spin and charge fluctuations on these bands.
They obtained a superconducting state without robust chiral edge modes 
along the boundary, thereby explaining the absence of experimentally detectable edge currents. 
Based on the Coulomb interaction mechanism presented in \S~\ref{theorySec}, 
Tada \textit{et al}. showed that the TRS 
can be restored near the (001) surface because of a Rashba-type spin-orbit interaction. \cite{Tada2009NJP}
They suggested that such pairing state at the interface is a promising candidate 
for the recently proposed time-reversal invariant topological superconductivity carrying helical edge current. \cite{Tanaka2009PRB, Sato2009PRB}
The surface states they propose naturally lead to cancelled charge current without magnetic field around the edge,
although the authors did not explicitly mention a relation to the above experimental results.

These null results for edge currents may also be related to a long-standing profound question 
concerning the size of the ``intrinsic angular momentum'' of chiral $p$-wave superfluid. 
The chiral edge current originates from a topological property of the bulk and is related to the angular momentum of Cooper pairs.\cite{Furusaki2001PRB}
In a superconductor, Meissner screening current is spontaneously generated in response to the topological edge current.
For $^3$He-A phase with $\Delta(\mib{k})=\Delta_\mathrm{A}(k_x \pm {\rm i}k_y)$, the expected topological edge mass current has never been observed. 
The magnitude of the mass current depends on the magnitude of the total angular momentum $\langle L \rangle = (N \hbar/2)(\Delta/E_\mathrm{F})^\alpha$. 
Depending on whether all the fermions contribute their pair angular momenta or only those in the interaction shell contribute them, 
the value of $\alpha$ varies as $\alpha = 0$, 1 or 2. 
As Leggett notes, this problem ``is more than 30 years old and still has apparently not attained a universally agreed resolution.'' \cite{Leggett2006QL}
The estimated magnitude of the chiral edge charge current in \sro\ has been largely based on the assumption of $\alpha = 0$, 
but a reduction factor with $(\Delta/E_\mathrm{F}) \sim 10^{-3}$ may have to be included.

\subsection{Superconducting gap structure}
\label{SCgapSec}

At the time when the previous review by Mackenzie and Maeno was written, 
the main issue concerning the gap structure was apparent controversy 
between the observed low-lying quasi-particle excitations and the predicted fully-gapped state. 
The former was deduced from the $T$-linear electronic specific heat coefficient \cite{NishiZaki2000} $C/T \sim T$ and 
the $T$-cubed nuclear-spin relaxation rate \cite{Ishida2000PRL} $1/T_1 \sim T^3$, 
as well as from the thermal conductivity \cite{Tanatar2001PRB, Izawa2001PRL}, 
penetration-depth \cite{Bonalde2000PRL}, and ultrasound attenuation \cite{Lupien2001PRL}. 
The anticipation of the full gap is a consequence of the spin-triplet pairing state on the quasi-two dimensional Fermi surface in the tetragonal symmetry, 
$\Vec{d}(\mib{k})=\hat{z}\Delta_0(k_x + {\rm i}k_y)$ with the isotropic full gap $|\Delta(k_x, k_y)| = \Delta_0$. 
It was then recognized that multiband effects are important: 
a large superconducting gap corresponding to $\Tc$ opens up on the ``active'' $\gamma$ Fermi surface 
derived mainly from the $d_{xy}$ orbitals of the Ru-$4d$ electrons and having 57\% of the total electronic DOS, 
while the ``passive'' $\alpha$ and $\beta$ Fermi surfaces derived mainly from the $d_{yz}$ and $d_{zx}$ orbitals host 
a gap substantially smaller compared with the energy scale of $\Tc$. 
The plateau in $C/T$ with increasing $H_{\parallel ab}$ is ascribable to the presence of the small gap 
in the passive bands with the DOS amounting to 43\%. 
Moreover, nearly-zero gap is expected for these passive bands. 
At the same time, various models have also been proposed which contain ``horizontal'' line nodes, 
with the gap zero at a certain $k_z$. 

A solution to these apparent controversies has been largely given by the specific heat 
under magnetic fields orientated accurately to the crystalline axes \cite{Deguchi2004PRL,Deguchi2004}. 
Before describing details of the superconducting gap structure investigation from these field-angle controlled specific-heat measurements, 
we first review the theoretically predicted orbital-dependent gap structure.
At present, the gap structure semi-quantitatively consistent with the existing experiments is 
the one predicted by Nomura and Yamada \cite{Nomura2002c}. 
As shown in Fig.~\ref{Fig3_5}, 
it consists of a large and anisotropic full-gap on the $\gamma$ Fermi surface, 
having a gap minimum along the $\Gamma$-M direction (corresponding to the Ru-O [100] direction), 
and of smaller gaps on the $\alpha$ and $\beta$ Fermi surfaces, amounting to 1/3 - 1/5 of the gap size on the $\gamma$ Fermi surface, 
and having a strong gap minimum accidentally almost zero along the $\Gamma$-X (diagonal) directions. 

The origins of such ``orbital-dependent'' gap structure are attributable to the differences 
in orbital symmetry of each band and in the resulting contribution to the pairing mechanism of each band. 
Such multiband gap structures with difference in size and anisotropy, introduced in connection with \sro \cite{Agterberg1997}, 
are proven to serve as an important concept also in the superconductivity of other systems 
such as MgB$_2$ \cite{Souma2003Nature} and iron-pnictides \cite{Kuroki2008PRL}. 

\begin{figure}
\begin{center}
\includegraphics[width=2.8in]{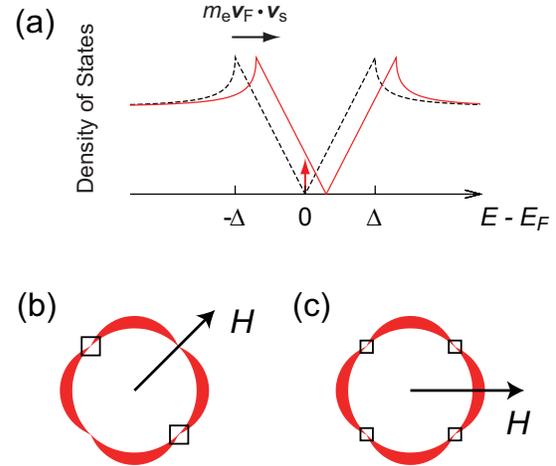}
\caption{
(a) Volovik effect: the energy shift of a quasi-particle spectrum due to the Doppler effect. 
Quasi-particle excitations around the nodes or minima of the superconducting gap for the applied magnetic fields 
parallel to (b) the nodal (or gap minimum) direction and (c) anti-nodal (or gap maximum) direction.
}
\label{Fig2_3_Volovik}
\end{center}
\end{figure}

We next describe somewhat in more detail how the anisotropic gap structure has been deduced from the specific heat \cite{Deguchi2004PRL}. 
The quasi-particle spectroscopy using the specific heat under oriented magnetic fields bases its principle on the so-called Volovik effect \cite{Volovik1993JETPL}. 
In the mixed state, the supercurrent velocities around the vortices are perpendicular to the applied field. 
Since the quasi-particle excitation spectrum is determined with respect to the rest frame of the condensates, 
the presence of the supercurrents results in the Doppler-shift in the quasi-particle spectrum in the laboratory frame 
as illustrated in Fig.~\ref{Fig2_3_Volovik}(a): 
$E(\mib{k}) = E_0(\mib{k}) + m_\mathrm{e} \mib{v}_\mathrm{F}(\mib{k}) \cdot \mib{v}_\mathrm{s}$, 
where $E(\mib{k})$ and $E_0(\mib{k})$ are the quasi-particle spectra in the frames of the laboratory and the condensate, 
$m_{\rm e}$ is the electron mass, $\mib{v}_\mathrm{F}$ and $\mib{v}_\mathrm{s}$ are the Fermi velocity and the supercurrent velocity, and 
$\mib{k}$ is the quasi-particle wave vector in the laboratory frame. 
For quasi-two-dimensional superconductors, 
$\mib{v}_\mathrm{F}$ is expected to be parallel to $\mib{k}$. 
As illustrated in Fig.~\ref{Fig2_3_Volovik}(b), this Doppler shift creates 
field-induced quasi-particle excitations in the regions of the $k$-vector around gap nodes or gap minima. 
Such field-induced quasi-particle excitations do not occur for the $k$-vector parallel to the applied field 
since $\mib{v}_{\rm F}$ is perpendicular to $\mib{v}_\mathrm{s}$. 
For systems with four line nodes on the two-dimensional Fermi surface, for example, 
the total quasi-particle excitations are minimized for the field direction along the nodes. \cite{Vekhter1999PRB}
Since the specific heat divided by $T$ or thermal conductivity divided by $T$ is proportional to the quasi-particle DOS at low temperatures, 
it takes a minimum if the field points to the direction of a nodal or minimum gap position. 

\begin{figure}
\begin{center}
\includegraphics[width=3in]{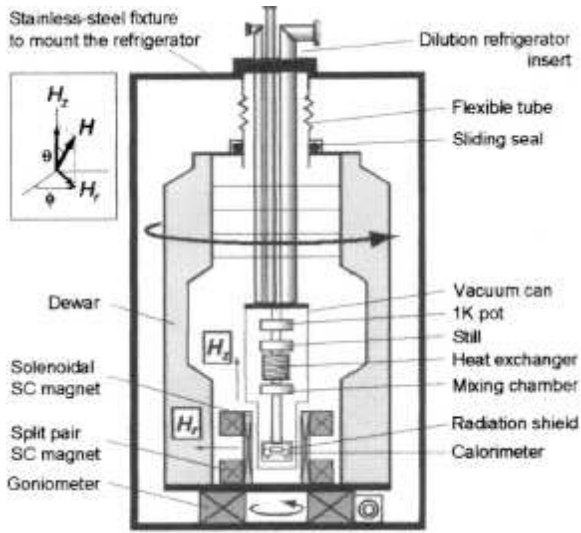}
\caption{
Apparatus used to measure anisotropic superconducting properties 
under magnetic fields of accurately and precisely aligned directions. 
From Deguchi \textit{et al}. (2004).\cite{Deguchi2004RSI}
}
\label{2-3_vecmag}
\end{center}
\end{figure}

An apparatus developed to investigate the field-angle quasi-particle spectra is depicted in Fig.~\ref{2-3_vecmag} \cite{Deguchi2004RSI}. 
In this apparatus, the so-called ``vector magnet'' consisting of split-pair superconducting magnet 
providing the horizontal field and the solenoid magnet providing the vertical field is placed in the dewar. 
This dewar rotates horizontally on the platform while the dilution refrigerator inserted in the dewar is fixed. 
This apparatus allows field alignment with respect to the crystalline axes of the sample with high accuracy and precision: 
the precision in the azimuthal angle of $\Delta\phi \sim 0.001^\circ$ and that of the polar angle of $\Delta\theta \sim 0.01^\circ$ are readily achieved. 
Because we do not need to rotate the refrigerator, the thermal stability of the measurements is greatly enhanced. 

\begin{figure}
\begin{center}
\includegraphics[width=3.2in]{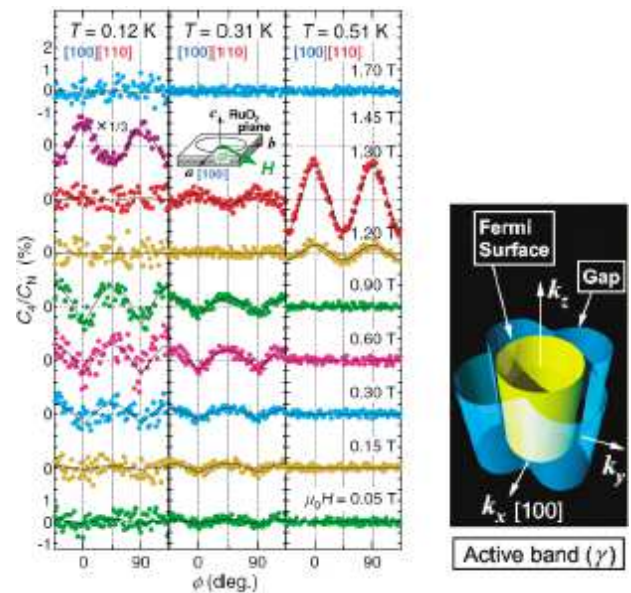}
\caption{
(Color online) (left) Fourfold oscillation of specific heat of \sro\ under field rotation within the basal plane. 
Oscillations at high fields are attributable to the anisotropy of $\Hcc$. \cite{Deguchi2004}
In the low-temperature and low-field region, the oscillation emerges with the opposite phase from the $\Hcc$ anisotropy. 
(right) Superconducting gap anisotropy on the active $\gamma$ band, deduced from the specific-heat oscillations. 
The gap minima are in the [100] directions.
From Deguchi \textit{et al}. (2004).\cite{Deguchi2004PRL}
}
\label{2-3_angularCp}
\end{center}
\end{figure}

Figure~\ref{2-3_angularCp} represents $C/T$ with the magnetic field rotating in the basal plane. 
At relatively high temperatures and near $\Hcc$, 
the four-fold oscillatory component of $C/T$ with minimum along the [110] direction is governed 
by the anisotropy of $\Hcc$ with the maximum in that direction below 0.8~K. \cite{Kusunose2004JPSJ,Udagawa2004PRB}
The data at 0.31~K and 0.12~K clearly exhibit a reversal of the $C/T$ oscillation when the field is reduced away from $\Hcc$. 
In the Doppler shift analysis, the minimum along the [100] direction in this low-$H$ and low-$T$ region indicates that 
the minimum in the main superconducting gap is in the [100] direction.

\begin{figure}
\begin{center}
\includegraphics[width=3.2in]{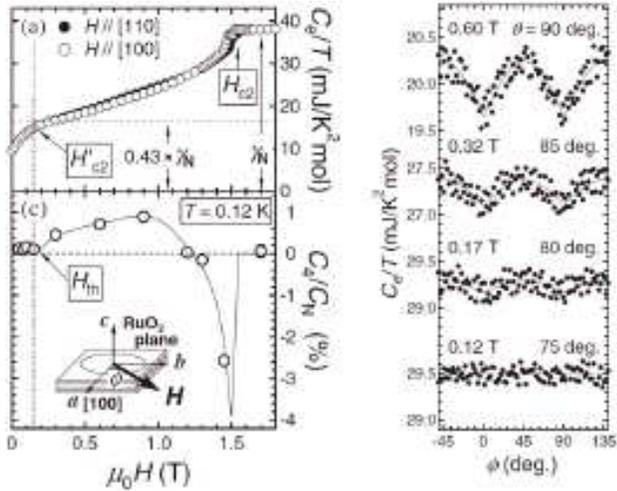}
\caption{
(left) Correspondence between the quasi-particle excitations under magnetic fields precisely aligned 
within the basal plane and the magnitudes of the fourfold oscillation amplitude of the specific heat. 
(right) Dependence on the azimuthal angle $\phi$ of the specific heat at $T = 0.12$~K 
for the same reduce field strength of $H/\Hcc(T)$. 
The disappearance of the oscillation at lower fields is attributable to the competition 
between the directions of the superconducting gap anisotropy 
in the active ($\gamma$) and passive ($\alpha$ and $\beta$) bands. 
From Deguchi \textit{et al}. (2004).\cite{Deguchi2004}
}
\label{2-3_angularCp2}
\end{center}
\end{figure}

Curiously, as the field is reduced below a certain value at the lowest measurement temperature, 
the oscillatory amplitude in $C/T$ diminishes (Fig.~\ref{2-3_angularCp2}), contrary to the expectation of growing relative amplitude for a single-band case. 
Since the threshold field of diminishing oscillation corresponds well with the exhaustion of the quasi-particle excitations 
corresponding to the passive $\alpha$ and $\beta$ bands, 
it is most naturally explained by the competition of the oscillations 
between the active and passive bands with opposite gap anisotropies. 
The specific-heat data under fields tilted from the basal plane assures this interpretation \cite{Deguchi2004}.

In addition to the anisotropies due to $\Hcc$ and nodal gap structure, 
recent experimental and theoretical studies revealed that 
the in-plane anisotropy of the specific heat or thermal conductivity in nodal superconductors can 
change sign, depending on the temperature and field range, due to the vortex scattering effect under rotating magnetic field \cite{Vorontsov2006PRL}. 
In fact, such a sign change was observed at $T \simeq 0.1\Tc$ in $C/T$ of the $d_{x^2-y^2}$-wave superconductor CeCoIn$_5$ \cite{An2010PRL}. 
For \sro, the observed large anisotropy at low temperatures probably represents 
the gap anisotropy of the main band since the measurements were extended to below $0.1\Tc$, 
and the reduction of the anisotropy at the lowest temperature of the measurements is attributed to the multiband effect.

The gap anisotropy consistent with these observations has been predicted by 
the microscopic theory based on the three-band Hubbard model by Nomura and Yamada \cite{Nomura2002c}. 
According to their third-order perturbation theory, described in some detail in \S~3, 
vertex correction terms beyond the second-order spin fluctuation terms play the essential roles for the pairing; 
the effects of Coulomb repulsion beyond the boson exchange provide the mechanism of unconventional superconductivity. 
In the heavily doped case as in \sro\ with four electrons with degenerate three bands, 
a $p$-wave state is more stable than the $d$-wave states. 
Odd parity associated with the $p$-wave symmetry requires the zero-gap at the symmetry points at the Brillouin-zone (BZ) boundary; 
naturally the main superconducting gap takes a minimum at the Fermi surface M-point, because it is closest to the BZ symmetry point. 
The anti-phase anisotropy on the passive bands is also a natural consequence 
if the antiferromagnetic spin fluctuations due to the nesting of $\alpha$ and $\beta$ bands are not favorable to the pairing. 
Thus, the essential superconducting symmetry is given by the chiral $p$-wave $\Vec{d}(\mib{k})=\hat{z}\Delta_0(k_x \pm {\rm i}k_y)$, 
while the low-lying quasi-particle excitations are attributable to the nearly-zero gap on the passive bands forced by the nesting. 
We will review other theoretical developments in \S~3.

\subsection{$d$-vector state}
\label{dvecSec}

Theoretically, several contributions to determine the direction of the $d$-vector have been considered. 
The determining mechanisms of the $d$-vector orientation can be divided into two categories \cite{Ishida2008JPCS}. 
One is of the ``atomic'' origin: orbital wave functions of Ru-$4d$ and O-$2p$ affect the anisotropy of the spins 
in both the normal and superconducting states. 
Such atomic spin-orbit interaction (SOI) energy is estimated to be about 50 meV \cite{Yanase2003b}. 
(More recently, the SOI energy is calculated to be 90 meV at the $\Gamma$ point by nonrelativistic band calculations \cite{Haverkort2008PRL},
and to be at least 100 meV based on the comparison between de Haas-van Alphen experiments and 
relativistic band calculations with local density approximation (LDA) \cite{Rozbicki2011JPCM}.)
Such SOI energy needs to be considered for electron Cooper pairs, while irrelevant to atomic superfluidity of $^3$He.
In the normal state, the anisotropy of the susceptibility is described as $\chi_c(T) = \chi_0 + 1.02 \chi_{ab}(T)$, 
with the temperature-independent part $\chi_0$ amounting to about 4\% of the total susceptibility. \cite{Ishida1997PRB}
The anisotropy of the temperature-dependent part, interpreted as the anisotropy of the spin susceptibility, is only about 2\%. 
Thus in the normal state, there is no strongly preferred orientation of the spins. 
In the superconducting state, the lifting of the degeneracies of the five possible $d$-vector states due to SOI decides 
the preferred orientation of the $d$-vector. 
From the microscopic theories of the pairing reviewed in \S~3, 
the magnitude of the splitting is expected to be four orders of magnitude smaller than SOI itself. 

The other mechanism to align the $d$-vector is due to dipolar interaction 
between the magnetic moments of the spins of a Cooper pair in the chiral state, 
which favors the $d$-vector parallel to the $L$-vector, namely along the $c$ direction. 
This mechanism dominates the determination of the $d$-vector orientation 
with respect to the orbital moment in chiral superfluid phase $^3$He-A. 
The manipulation of the $d$-vector orientation has been demonstrated by Ahonen \textit{et al}. \cite{Ahonen1976JLTP} 
For $^3$He confined within narrow parallel plates, 
the orbital motion is restricted and it is expected that the $L$-vector aligns perpendicular to the plates. 
The $d$-vector points in the same direction due to the dipolar interaction of about 5~mT. 
By the application of the in-plane field, the $d$-vector is indeed reoriented. 

The first mechanism, atomic SOI associated with the Ru-$4d$ orbitals, 
is incorporated in the microscopic pairing mechanism by Yanase and Ogata \cite{Yanase2003b}, 
yielding the $d$-vector pointing in the $c$ direction with the anisotropy energy corresponding to 90~mT. 
Later, Yoshioka and Miyake introduce the importance of the O-$2p$ orbitals in the Hubbard model (the $d$-$p$ model) and 
reproduced the $\Vec{d} \parallel c$ result, 
but at the same time show that $\Vec{d} \parallel ab$ becomes more stable 
if the on-site Coulomb repulsion at the oxygen site, $U_{pp}$, is comparable to $U_{dd}$. 
A more detailed discussion of the atomic SOI will be given in \S~\ref{RoleSOI}.
The second mechanism, the spin dipolar interaction within a Cooper pair, favors $\Vec{d} \parallel c$ 
with the energy corresponding to 22~mT. 
Unlike superfluid $^3$He, the charges of the Cooper-pair electrons lead to the orbital magnetic moment, 
associated with the orbital angular momentum $L$, which interacts with the spins. 
Such SOI within a Cooper pair yields the spins orienting parallel to $L$, 
thus $\Vec{d} \parallel ab$ by the energy corresponding to 64~mT \cite{Miyake2010JPSJ}. 
Within the five $d$-vector states allowed in the tetragonal crystal symmetry, 
a $d$-vector in the basal plane may be constructed as a mixing of four of the states. 
Whether such a state is allowed as the ground state is an issue of current debates. 
These proposed microscopic mechanisms of orienting the $d$-vector are summarized in Table~\ref{dvec}.

\begin{table*}[tb]
\begin{center}
\caption[]{
Various microscopic mechanisms of orienting the $d$-vector and the corresponding stability energies. 
The stability energies are expressed in units of magnetic field (mT), 
divided by the Bohr magneton $\mu_B$. 
$^{\ast)}$For $U_{pp} \gtrsim 3t_{dp}$. See \S~\ref{RoleSOI} for details concerning this condition for
the oxygen-site Coulomb interaction $U_{pp}$ and the hopping parameter $t_{dp}$. 
} 
\label{dvec}
\begin{tabular}{llllll}\hline
Interaction & Preferred orientation & Stability energy &\\ \hline
Atomic spin-orbit interaction (Ru site) & $\Vec{d} \parallel \Vec{L}$ ($\Vec{d} \parallel c$) & $\sim 90$~mT & (Ref.~\ref{Yanase2003a}) \\
Atomic spin-orbit interaction (O site) & $\Vec{d} \perp \Vec{L}$ ($\Vec{d} \parallel ab$) &  $\sim 50$~mT~$^{\ast)}$ & (Ref.~\ref{Yoshioka2009}) \\
Spin-dipolar interaction within a Cooper pair & $\Vec{d} \parallel \Vec{L}$ & $\sim 22$~mT & (Ref.~\ref{Hasegawa2003}) 
\\Spin-orbit interaction within a Cooper pair & $\Vec{d} \perp \Vec{L}$ & $\sim 64$~mT & (Ref.~\ref{Miyake2010JPSJ}) \\ 
\\
Zeeman interaction ($H \parallel c$) & $\Vec{d} \perp \Vec{L}$ & \\ 
Zeeman interaction ($H \parallel ab$) & $\Vec{d} \parallel \Vec{L}$ & \\ \hline
\end{tabular}
\end{center}
\end{table*}

With these competing mechanisms of comparable magnitudes, it is not easy to predict the $d$-vector orientation theoretically. 
We note that if the $d$-vector lies in the basal plane, as illustrated in Fig.~\ref{2-1_d-vector}(c), 
the pinning within the basal plane must be extremely weak because of the tetragonal symmetry. 
It would then be possible that the $d$-vector orients perpendicular to the external field of any directions; 
the spin susceptibility is unchanged across $\Tc$, consistent with the NMR Knight-shift observations 
for both $H \parallel ab$ and $H \parallel c$. 

\begin{figure}
\begin{center}
\includegraphics[width=3.2in]{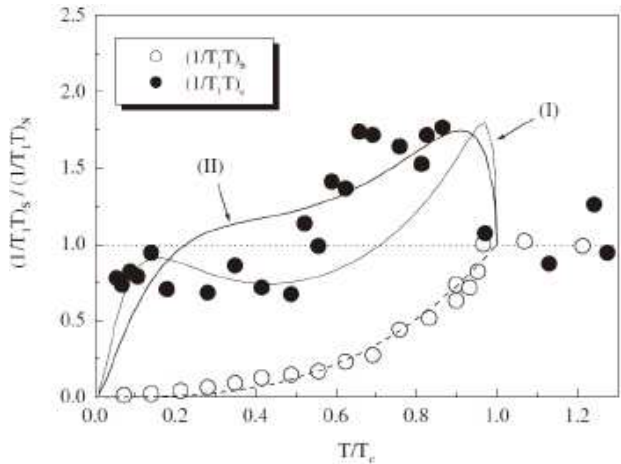}
\caption{
Spin-lattice relaxation rate of nuclear quadrupole resonance (NQR) of oxygen nuclei. 
Energy dissipations for both in-plane (open circles) and out-of-plane (closed circles) components 
in these zero-field data have been deduced. 
The anomalous enhancement of the out-of-plane absorption is attributed to the collective-mode excitation 
associated with the $d$-vector dynamics. \cite{Mukuda2002PRB}
The solid curves are theoretical analyses by Miyake.\cite{Miyake2010JPSJ}.
}
\label{2-3_T1}
\end{center}
\end{figure}

Collective excitations of Cooper pairs manifest the available internal spin and orbital degrees of freedom and 
give invaluable information of the pairing state. 
A curious anomaly of the spin-lattice relaxation rate of $^{17}$O 
under zero magnetic field (NQR) is present in the in-plane oxygen site. 
Since the principal axis for the oxygen quadrupole moment lies in the basal plane, 
$1/T_1$ consists of the spin fluctuations of both the in-plane and out-of-plane components. 
By comparing with the $^{101}$Ru NQR relaxation rate, Mukuda \textit{et al}. deduced the anisotropy in $1/T_1$ of $^{17}$O NQR, 
which exhibits a broad peak below $\Tc$. \cite{Mukuda2002PRB}
Miyake recently \cite{Miyake2010JPSJ} gives explanation of this behavior as 
due to the internal Josephson oscillation of the Cooper pairs. 
As shown in Fig.~\ref{2-3_T1}, 
the oscillation of the $d$-vector in the basal plane, 
corresponding to the fluctuation of Cooper-pair spins along the $c$ axis, 
leads to $1/T_1$ with characteristic two broad peaks below $\Tc$. 
Although not available at present, $^{17}$O NQR signals from the apical oxygen site would give confirming results for this interpretation.

A certain number of theoretical studies on possible collective modes 
have been done for the spin-triplet pairing state of Sr$_2$RuO$_4$.\cite{Tewordt1999,Tewordt1999-2,Kee2000,Kee2000-2,Fay2000,Hirashima2007,Nomura2008JPSJ}
Among possible collective modes, the most important one is a spin-wave 
mode which is related to fluctuations of the $d$-vector direction 
from its stable direction. 
If the anisotropy in spin space is neglected, then this mode corresponds 
to the Goldstone mode accompanied by the symmetry breaking 
with respect to rotations in spin space and its excitation energy equals 
zero at the long-wavelength limit $\mib{q}=0$ 
(because the $d$-vector rotates freely without any energy cost). 
In reality, this mode should have an energy gap at $\mib{q}=0$ 
(maybe not zero but small gap for Sr$_2$RuO$_4$), 
reflecting the anisotropy of the $d$-vector. 
According to the theoretical studies~\cite{Tewordt1999,Tewordt1999-2,Kee2000,Kee2000-2,Fay2000,Hirashima2007,Nomura2008JPSJ}, 
this mode may be observed experimentally 
in the in-plane dynamical spin susceptibility, for the chiral pairing state 
$\mib{d}(\mib{k}) \propto \hat{z}(k_x \pm {\rm i} k_y)$. 
Although such a collective spin excitation has not been observed experimentally, 
it might be observed in the future, using low-energy neutron scattering 
or other spin-resonance measurements. 
Discovery of such a spin collective mode could be strong evidence 
for the triplet pairing in Sr$_2$RuO$_4$.


\section{Pairing by Coulomb repulsion}
\label{theorySec}
\subsection{Microscopic pairing mechanisms based on Coulomb repulsion}

From theoretical points of view, many studies on the triplet pairing 
mechanism in Sr$_2$RuO$_4$ have been performed 
so far~\cite{Mazin1997,Mazin1999,Monthoux1999,Monthoux2005,
Kuzmin2000,Kuwabara2000,Sato2000,Kuroki2001,Ogata2002,Takimoto2000,
Baskaran1996,Spalek2001,Wysokinski2003,Koikegami2003,
Nomura2000a,Nomura2002a,Nomura2002b,Honerkamp2003,
Werner2003,Arita2004,Hoshihara2005,Raghu2010}. 
To our knowledge, all of them essentially require unconventional 
(i.e., non-phonon) mechanisms for explaining the triplet pairing. 
It is believed that strong electron correlations are much likely 
to prohibit conventional phonon-mediated pairing in Sr$_2$RuO$_4$. 
In fact, Sr$_2$RuO$_4$ is a typical strongly correlated electron system, 
as suggested from many experimental and theoretical studies. 
According to band-structure calculations~\cite{Oguchi1995, Singh1995}, 
Ru-$4d\varepsilon$ electrons, which are relatively localized, 
occupy the states near the Fermi level dominantly, 
and therefore the electronic properties will be affected significantly 
by electron correlations. Consistent with this, sizable mass enhancement 
was observed by quantum oscillation measurements~\cite{Mackenzie1996}. 
In addition, it was confirmed that a related compound Ca$_2$RuO$_4$ 
is a Mott insulator~\cite{Nakatsuji1997, Nakatsuji2000}. 

Here we overview various proposals on the unconventional pairing 
mechanism for Sr$_2$RuO$_4$. 
Most of proposed mechanisms can be classified, 
depending on the microscopic origin of the pairing interaction assumed: 
(i) ferromagnetic spin fluctuations or paramagnons 
as in the superfluid $^3$He~\cite{Mazin1997,Mazin1999,
Monthoux1999,Monthoux2005,Kuzmin2000,Hoshihara2005}, 
(ii) anisotropic incommensurate antiferromagnetic 
spin-fluctuations~\cite{Kuwabara2000,Sato2000,Kuroki2001,Ogata2002}, 
(iii) incommensurate charge or orbital fluctuations~\cite{Takimoto2000}, 
(iv) Hund's-rule coupling among Ru-$4d\varepsilon$ 
electrons~\cite{Baskaran1996,Spalek2001},  
(v) inter-site interaction (including inter-site 
Coulomb repulsion)~\cite{Wysokinski2003,Koikegami2003}, 
and (vi) on-site Coulomb repulsion between Ru-$4d\varepsilon$ electrons 
(particularly among Ru-$4d_{xy}$ electrons)~\cite{Nomura2000a,Nomura2002a,
Nomura2002b,Honerkamp2003}. 
At a first glance, (i) may seem the most plausible, 
since Sr$_2$RuO$_4$ has Fermi liquid parameters similar to those 
of superfluid $^3$He and the three-dimensional SrRuO$_3$ 
is a ferromagnet~\cite{Rice1995,Rice1998}. 
However, neutron scattering experiments revealed that the dominant magnetic 
correlation in Sr$_2$RuO$_4$ is incommensurate antiferromagnetic 
rather than ferromagnetic~\cite{Sidis1999}, 
reflecting the nesting character between the $\alpha$ 
and $\beta$ Fermi surfaces\cite{Mazin1999, Nomura2000b}. 
In (ii), this incommensurate fluctuation is assumed to be much anisotropic: 
$\chi_c({\mib Q})/\chi_{ab}({\mib Q}) > 4$ - 7 
(${\mib Q}$ is the incommensurate nesting vector)~\cite{Kuwabara2000,Kuroki2001}. 
Qualitatively, such anisotropy is indeed consistent with an NMR measurement~\cite{Ishida2001} 
and theoretical calculation~\cite{Eremin2002}, but seems too large 
($\chi_c({\mib Q})/\chi_{ab}({\mib Q}) \approx 3$ in the NMR experiment). 
In (iii), incommensurate charge fluctuations due to the nesting 
between the $\alpha$ and $\beta$ bands mediate the triplet pairing. 
In this mechanism, inter-orbital Coulomb interaction should 
be larger than the intra-orbital Coulomb repulsion somehow~\cite{Takimoto2000}. 
In the scenarios (ii) and (iii), the pairing interaction will become the strongest 
on the $\alpha$ and $\beta$ bands rather than on the main band $\gamma$, 
and therefore the $\alpha$ and $\beta$ bands will be active ones 
in pairing transition rather than the $\gamma$ band, 
contradicting several experimental results which suggest 
the dominance of the $\gamma$ band. 
In (iv), the Hund's-rule coupling is believed to stabilize 
the parallel spins of triplet Cooper pairs~\cite{Baskaran1996,Spalek2001}. 
However, the Hund's coupling between the local Ru-$4d$ orbitals can stabilize parallel 
spin states only on each local Ru site, but does not necessarily stabilize them 
over the superconducting coherence length. 
It is still unclear whether the Hund's-rule coupling stabilizes 
spin-triplet pairing rather than spin-singlet pairing in Sr$_2$RuO$_4$ or not. 
In (v), as easily shown, the in-plane inter-site attractive interaction 
between nearest-neighbor sites induces 
$\sin k_x$ and $\sin k_y$-wave pairing~\cite{Wysokinski2003}. 
As for the gap structure in (v), more inter-site couplings 
between farther sites become necessary 
to explain the higher-order harmonics of superconducting gap function. 
If the inter-layer interactions between two sites 
on adjacent different RuO$_2$ layers are included, 
then horizontal line node appears 
in the superconducting gap~\cite{Hasegawa2000,Zhitomirsky2001,
Wysokinski2003,Koikegami2003}. 
However, such a horizontal node seems inconsistent 
with a result of field-oriented specific heat measurement~\cite{Deguchi2004}. 

Recently, Raghu \textit{et al}. presented a two-band calculation 
for the one-dimensional $yz$ and $xz$ bands, 
in order to explain the absence of edge currents~\cite{Raghu2010}. 
For pairing interaction, they used the expression of the second-order perturbation 
in the on-site intra- and inter-orbital repulsion. 
This theory seems substantially the same as in (iii). 
In fact, the obtained momentum dependence of the gap function 
is very similar to that obtained in the random phase approximation (RPA) calculation of Ref.~\ref{Takimoto2000}, 
i.e., the momentum dependence is well approximated by the forms 
$\Delta_{xz}({\mib k}) \approx \Delta_0\sin k_x \cos k_y$ 
and $\Delta_{yz}({\mib k}) \approx \Delta_0\cos k_x \sin k_y$ in the both theories. 
The both theories suggest that the triplet pairing becomes more favorable 
against the singlet pairings, as the inter-orbital repulsion becomes stronger. 

At present, we consider that (vi) is the most plausible scenario, 
in that it adopts only simple assumptions whose origins are microscopically clear 
or which are consistent with most of experiments. 
In the next section we will describe this mechanism in detail. 

\subsection{Single-band Hubbard model}

To describe the above mentioned mechanism (vi), 
we use weak-coupling approximation, 
but require nothing except a realistic tight-binding electronic structure 
and the on-site Coulomb repulsion at each Ru site. 
First, we consider only the $\gamma$ band for simplicity, 
since the $\gamma$ band takes the major part of the density of states 
(about 57\%) and the heaviest effective mass 
(14.6 times the bare-electron mass) 
at the Fermi level~\cite{Mackenzie1996}. 
We describe the $\gamma$ band by a single-band $t$-$t'$ 
Hubbard model on a two-dimensional square lattice~\cite{Nomura2000a}: 
\begin{equation}
H = t \sum_{\langle i,j \rangle, \sigma}^{\rm n.n.} a_{i\sigma}^{\dag} a_{j\sigma} 
+ t' \sum_{\langle i,j \rangle, \sigma}^{\rm n.n.n.} a_{i\sigma}^{\dag} a_{j\sigma} 
+ U \sum_i a_{i\uparrow}^{\dag} a_{i\downarrow}^{\dag}
a_{i\downarrow} a_{i\uparrow}, 
\end{equation}
where $a_{i \sigma}$ ($a_{i \sigma}^{\dag}$) is the annihilation (creation) 
operator of an electron with spin $\sigma$ at Ru site $i$. 
Here $t$ and $t'$ are the hopping integrals between nearest and next-nearest 
neighbor sites respectively, $U$ is the on-site Coulomb integral. 
We take $t=-1$, $t'=-0.375$, which can reproduce the electronic structure 
of the $\gamma$ band approximately. 
Roughly speaking, since the $\gamma$ band is constructed mainly 
from the Ru-$4d_{xy}$ orbitals, the microscopic origin of $U$ is 
the Coulomb repulsion between the Ru-$4d_{xy}$ electrons. 
$t$ is estimated to be about 0.25 eV, if the bare bandwidth equals 2 eV. 
``n.n.'' (``n.n.n.'') means that the summation with respect to $i$ and $j$
is restricted to nearest (next-nearest) neighbor sites. 
The superconducting transition point is determined 
by solving the Eliashberg equation~\cite{Nomura2000a}: 
\begin{equation}
\Delta_{\sigma_1\sigma_2}(k) = - \frac{T}{N} \sum_{k'\sigma_3\sigma_4} 
V_{\sigma_1\sigma_2,\sigma_4\sigma_3}(k, k') |G(k')|^2 \Delta_{\sigma_3\sigma_4}(k'), 
\label{Eq:Eliashberg}
\end{equation}
where $k=({\mib k}, i \omega$), and $\Delta_{\sigma\sigma'}(k)$ 
is the superconducting order parameter (in other words, the wave function of Cooper pairs). 
$G(k)$ is the particle propagator for quasi-particles. 
$V_{\sigma_1\sigma_2,\sigma_4\sigma_3}(k, k')$ is the pairing interaction, in other words, 
the scattering amplitude between pairing quasi-particles, 
where the momenta and spins of the quasi-particles 
in the initial (final) state are ${\mib k}' \sigma_3$ and $-{\mib k}' \sigma_4$ 
(${\mib k} \sigma_1$ and $-{\mib k} \sigma_2$). 
Transition temperature is the highest temperature below 
which a non-trivial solution $\Delta_{\sigma\sigma'}(k) \neq 0$ appears, 
and the pairing symmetry is determined by the momentum and spin dependences 
of the solution $\Delta_{\sigma\sigma'}(k)$. 

In the weak-coupling regime, we can evaluate the pairing interaction 
by perturbation expansion in $U$: 
\begin{equation}
V(k, k') = U V^{(1)}(k,k') + U^2 V^{(2)}(k, k') + U^3 V^{(3)}(k, k') + \cdots. 
\label{Eq:Expansion}
\end{equation}
Here we have omitted the spin indices for simplicity. 
The first-order term remains non-zero only in the cases of singlet pairing. 
This first-order term is independent of momentum and frequency, 
and is repulsive ($U V^{(1)}(k, k')=U > 0$) in the $s$-wave channel. 
This is the reason why the $s$-wave pairing is generally prohibited 
in strongly correlated electron systems. 
Higher-order terms can possess non-trivial momentum dependences, 
as a result from many-body correlation. 
The possibility of anisotropic (i.e., non-$s$-wave) pairing depends 
on whether this non-trivial momentum dependence 
is attractive for the corresponding anisotropic pairing channel or not.  
Thus we turn our attention to the higher-order terms. 

\begin{figure}
\begin{center}
\includegraphics[width=3.2in]{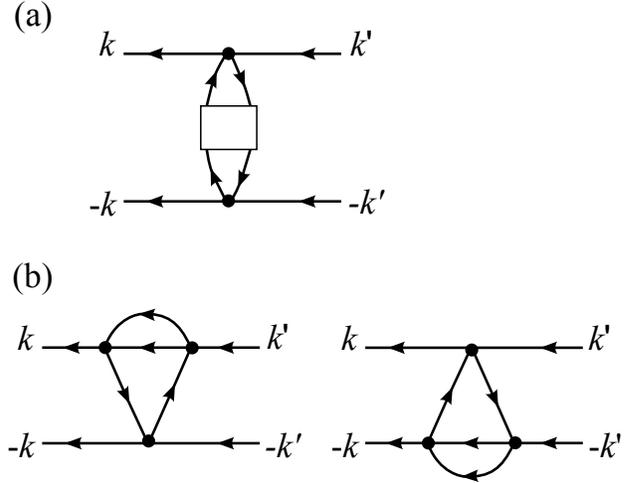}
\end{center}
\caption{
Diagrammatic representations of typical pair-scattering processes. 
The filled four-leg vertices and oriented solid lines denote the on-site repulsion 
$U$ and particle propagator, respectively. 
(a) General diagrammatic representation of the pair scattering processes 
which can be regarded as due to exchange of elementary excitations. 
The rectangle represents arbitrary diagrammatic structure 
connecting upper and lower parts. 
Diagrams of this type provide contributions depending only 
on momentum-energy transfer $q=k-k'$ for the pairing interaction. 
(b) Typical third-order vertex-corrected terms. 
Note that these processes are not included in the processes of the above type (a). 
The diagrams in (b) play an essential role for triplet $p$-wave pairing.}
\label{Fig3_1}
\end{figure}

The second-order contribution is expressed using the bare susceptibility 
$\chi_0(q)$ as $U^2 V^{(2)}(k, k') = \pm U^2 \chi_0(k-k')$, 
where the sign is plus (minus) for singlet (triplet) pairing. 
In many cases, the bare susceptibility $\chi_0(q)$ captures qualitatively 
the correct momentum dependence of more precise susceptibility $\chi(q)$ 
calculated within advanced approximations, e.g., RPA~\cite{Moriya1985}, 
fluctuation-exchange approximation (FLEX)~\cite{Bickers1989}, etc., 
although the momentum and frequency dependences of the accurate $\chi(q)$ 
become more enhanced and steeper near magnetic transition points 
than those of the bare $\chi_0(q)$. 
In pairing theories based on spin-fluctuation mediation, 
it is assumed that the pairing interaction can be expressed 
by the form $V(k, k') \propto \chi(k-k')$, where $\chi(q)$ is calculated 
by those approximate methods~\cite{Scalapino1986,Bickers1989}. 
Therefore, as far as we discuss qualitative features including the most favorable 
pairing symmetry, the second-order theory can often predict qualitatively 
the same results as other methods based on spin-fluctuation mediation. 
These theories present a clear physical picture that a pair of quasi-particles 
are formed into a Cooper pair by exchanging elementary excitation 
(`spin fluctuation' in the present case), where $V(k, k')$ is a function 
only of the momentum-energy transfer $q=k-k'$, i.e., the momentum and energy 
of the elementary excitation. 
Pair-scattering processes due to exchange of elementary excitations 
are generally represented by such a type of diagrams as in Fig.~\ref{Fig3_1}(a). 

Next we proceed to the third-order theory. 
The third-order term $U^3 V^{(3)}(k, k')$ can bring about qualitatively new momentum 
dependence, as mentioned below. 
The third-order term consists of RPA-like terms 
and so-called ``vertex-corrected'' terms: 
$U^3 V^{(3)}(k, k')= U^3V^{({\rm RPA})}(k, k') + U^3V^{({\rm V.C.})}(k, k')$
(Analytic expressions of the third-order terms are presented in Ref.~\ref{Nomura2000a}). 
The RPA-like terms, which are included also within RPA, present qualitatively 
the same momentum dependence as the second-order term, 
and are expressed by diagrams of the type (a) of Fig.~\ref{Fig3_1}. 
On the other hand, the vertex-corrected terms do not depend only 
on $k-k'$, but on both $k$ and $k'$. 
Typical examples of third-order vertex-corrected terms are shown in Fig.~\ref{Fig3_1}(b). 
To be regarded as mediated by some kinds of elementary excitations of the system, 
$V(k, k')$ must be a function only of the momentum and energy ($q=k-k'$) 
of elementary excitations excited in the intermediate states. 
Therefore the vertex-corrected terms cannot be regarded as 
due to exchange processes of any elementary excitations. 

\begin{figure}
\begin{center}
\includegraphics[width=3.2in]{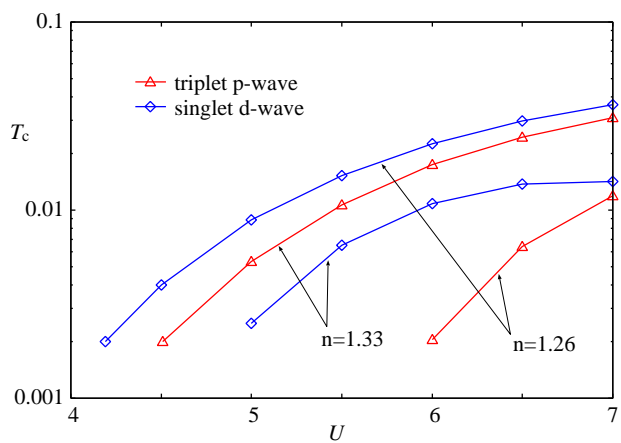}
\end{center}
\caption{(Color online) 
Transition temperature $T_{\rm c}$ as a function 
of the on-site repulsion $U$, calculated by using 
a single-band Hubbard model for the $\gamma$ band~\cite{Nomura2000a}. 
$n$ denotes the total electron filling ($n=1.33$ is the realistic value 
for the $\gamma$ band of Sr$_2$RuO$_4$).
Half filling is $n=1$.}
\label{Fig3_2}
\end{figure}

\begin{figure}
\begin{center}
\includegraphics[width=3.2in]{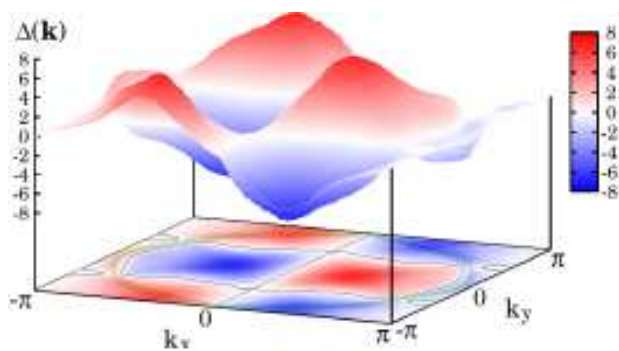}
\end{center}
\caption{(Color online) 
Momentum dependence of the triplet superconducting 
order parameter $\Delta(k)$ (only the $k_x$-type). 
The green circle and gray solid lines on the base plane represent 
the $\gamma$ Fermi surface and the zeros of $\Delta(k)$, respectively.}
\label{Fig3_3}
\end{figure}

Within the third-order perturbation theory, 
Nomura and Yamada  showed that the most probable pairing symmetry 
on the $\gamma$ band (Ru-$4d_{xy}$ band) is the spin-triplet $p$-wave~\cite{Nomura2000a}. 
In Fig.~\ref{Fig3_2}, transition temperatures calculated for singlet and triplet pairing states 
are displayed as a function of $U$. 
For the total electron filling of the $\gamma$ band $n=1.33$, the highest 
transition temperature is given by a triplet pairing state, 
while the singlet $d_{x^2-y^2}$-wave state becomes favorable toward half filling. 
Thus, the fact that the electron filling of the $\gamma$ band is away from half filling 
is essential for realization of triplet pairing in Sr$_2$RuO$_4$. 
The momentum dependence of the order parameter $\Delta(k)$ 
suggests the $p$-wave symmetry, as shown in Fig.~\ref{Fig3_3} 
(Only the $k_x$-type solution $\Delta_x(k)$ is displayed. 
The $k_y$-type solution $\Delta_y(k)$ is obtained by $\pi/2$ rotation 
around the $z$ axis. $\Delta_x(k)$ and $\Delta_y(k)$ transform as $k_x$ and $k_y$ 
under the $D_{4h}$ symmetry operations, respectively). 
Here note that the momentum dependence of  $\Delta_x(k)$ 
is not similar to the lowest-order harmonics $\sin k_x$, 
but seems to contain higher-order harmonics. 
Although the dominant harmonics is $\sin k_x \cos k_y$, the zero contour 
of $\Delta_x(k)$ crosses the Fermi circle only at two Fermi surface positions $k_x=0$. 
This suggests that line nodes occur at $k_x=0$ on the realistic cylindrical Fermi surface. 
However, it should be noted that these line nodes do not cause zero-gap 
in the energy spectrum, if we also consider $\Delta_y(k)$ of the $k_y$ type as well, 
because the real energy gap is given 
by $|\Delta({\mib k})|=[\Delta_x(k)^2+\Delta_y(k)^2]^{1/2}|_{\omega=0}$ 
within weak-coupling theories. 

\begin{figure}
\begin{center}
\includegraphics[width=3.2in]{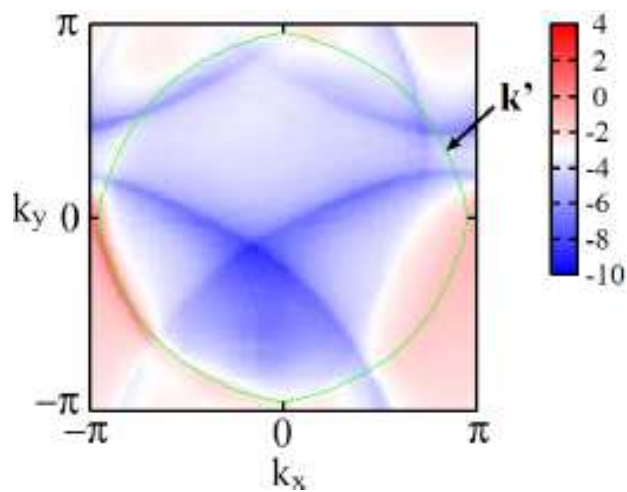}
\end{center}
\caption{(Color online) 
${\mib k}$-dependence of the triplet-pairing interaction $V(k,k')$. 
${\mib k}'$ is fixed at the point indicated by the arrow. 
The green circle represents the $\gamma$ Fermi surface. 
$U=4$ and $T=0.002$.}
\label{Fig3_4}
\end{figure}

In Fig.~\ref{Fig3_4}, $V(k, k')$ calculated within the third-order perturbation theory in $U$ 
is plotted as a function of $k$ for fixed $k'$. 
We pay attention to the behavior of $V(k, k')$ near the Fermi surface. 
We should note that $V(k, k')$ takes low values around ${\mib k}={\mib k}'$ 
and the maximum value around ${\mib k}=-{\mib k}'$. 
The low values of $V(k, k')$ around ${\mib k}={\mib k}'$ 
is attributed to the weak ferromagnetic feature 
of $\chi_0(q)$  (remember that the second-order contribution 
is expressed by $U^2 V^{(2)}(k, k')=-U^2\chi_0(k-k')$). 
This ferromagnetic feature is due to the fact that the Fermi level is close 
to the van Hove singularity on the $\gamma$ band~\cite{Nomura2000b}. 
On the other hand, the high values around ${\mib k}=-{\mib k}'$ are brought 
about by the third-order vertex-corrected term $U^3 V^{({\rm V.C.})}(k,k')$. 
As inferred from the minus sign of the right-hand side of the Eliashberg equation (\ref{Eq:Eliashberg}), 
this characteristic momentum dependence of $V(k, k')$ favors sign change of $\Delta(k)$ 
between ${\mib k}$ and $-{\mib k}$, and accordingly causes $p$-wave pairing. 
It is interesting that the same vertex-corrected terms as shown in Fig.~\ref{Fig3_1}(b) 
cause $p$-wave pairing also in two-dimensional isotropic repulsive fermion systems 
in the weak-coupling regime~\cite{Chubukov1993,Feldman1997}. 
At present, we do not know how to draw such a clear and intuitive physical picture 
on the momentum dependence derived from vertex-corrected terms 
as that on the momentum dependence due to elementary-excitation mediation. 
The third-order perturbation calculation has been applied 
to other unconventional superconductors than Sr$_2$RuO$_4$, 
high-$T_{\rm c}$ cuprates, organics, several heavy-fermion systems, 
and suggested a plausible pairing symmetry for each of them, so far~\cite{Yanase2003a}.  
Effects of higher-order contributions are discussed in the Appendix of Ref.~\ref{Yanase2003a}.

Honerkamp and Salmhofer found the $p$-wave superconducting phase 
by applying the one-loop renormalization-group (RG) theory 
to a similar Hubbard model~\cite{Honerkamp2001,Honerkamp2003}. 
Although the RG analysis does not include one-particle self-energy corrections, 
it has an advantage that the magnetic, pairing, and other instabilities 
can be treated on an equal footing. 
According to their result, the momentum dependence of the effective interaction 
seems similar to the one obtained within the above third-order perturbation theory, 
rather than to the one induced by the ferromagnetic spin fluctuation, 
in spite that the $p$-wave superconducting phase is adjacent to a ferromagnetic phase. 
In fact, the RG calculation seems to include the same vertex corrections 
as the third-order perturbation calculation. 

\subsection{Extension to three-band model} 

Nomura and Yamada extended the above single-band calculation 
to a three-band calculation~\cite{Nomura2002a,Nomura2002b}. 
They used a three-band Hubbard model (without SOI) 
to describe the realistic three-band electronic structure of Sr$_2$RuO$_4$. 
In the three-band calculation, three components of order parameter, 
$\Delta_a(k)$ ($a=\alpha, \beta, \gamma$), are required 
for describing pairing on each band. 
They determined $\Delta_a(k)$ by solving the Eliashberg equation 
within the third-order perturbation in the on-site Coulomb integrals. 
According to their result, the superconducting order parameter 
takes the maximum value on the $\gamma$ band. 
Thus the `orbital dependent superconductivity' initially suggested 
in Ref.~\ref{Agterberg1997} is verified by the microscopic theory. 
The probable pairing symmetry is reasonably the $p$-wave again, 
and the momentum dependence of $\Delta_{\gamma}(k) $ is almost 
the same as that by the above single-band 
calculation for the $\gamma$ band. 
According to their three-band calculation, the picture that 
inter-band proximity~\cite{Zhitomirsky2001} 
induces pairing on the passive $\alpha$ and $\beta$ bands is indeed valid. 
In fact, they found that the pairing amplitude $\Delta_{\gamma}(k)$ on the $\gamma$ band 
remains finite and $\Delta_{\alpha,\beta}(k)$ on the passive bands become zero, 
when the inter-orbital Coulomb interaction is artificially set to zero. 

\begin{figure}
\begin{center}
\includegraphics[width=3.2in]{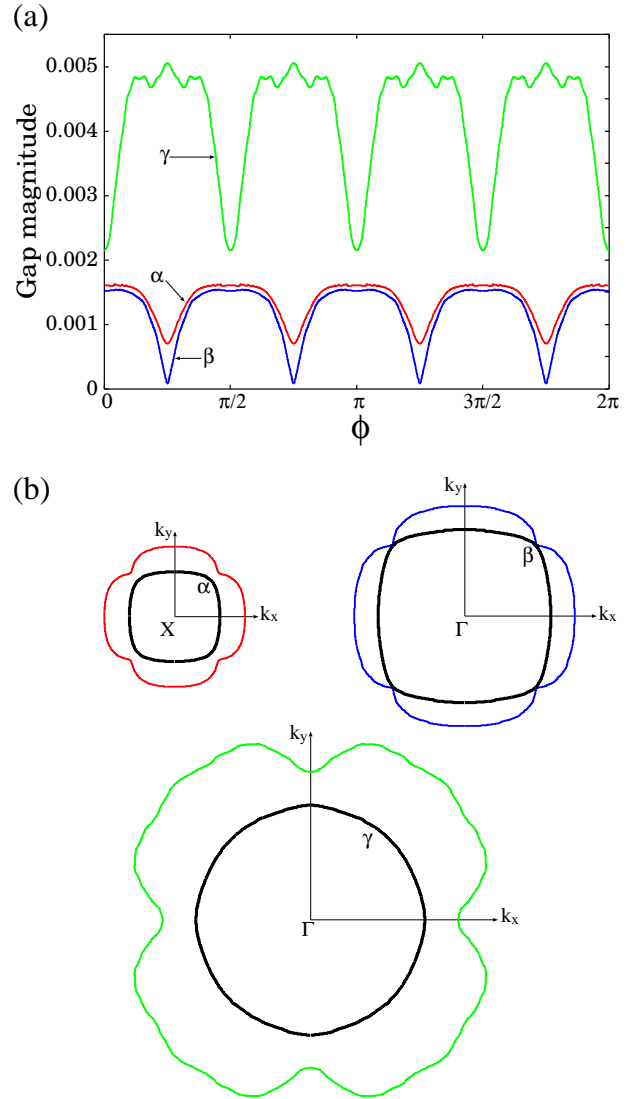}
\end{center}
\caption{(Color online) 
Predicted superconducting gap structure~\cite{Nomura2002c,Nomura2005}. 
(a)~Gap magnitude on each of the three Fermi surfaces 
is depicted as a function of the azimuthal angle $\phi$ around the $c$ axis 
(the $a$ axis direction corresponds to $\phi=0$). 
The unit of energy on the vertical axis is about 0.057 eV. 
(b)~The Fermi surfaces are represented by the solid black circles. 
The dependence of gap magnitude on in-plane direction is expressed 
by the distance from the Fermi circle along the direction. }
\label{Fig3_5}
\end{figure}

Nomura and Yamada used the three-band solution 
$\Delta_a(k)$ ($a=\alpha, \beta, \gamma$) 
for the gap function~\cite{Nomura2002c,Nomura2005}. 
If any of the five $d$-vector states displayed in Table~\ref{table3} is realized, 
the gap magnitude on each band is calculated by 
$|\Delta_a({\mib k})|=[{\Delta_{a,x}(k)}^2+{\Delta_{a,y}(k)}^2]^{1/2}|_{\omega=0}$, 
where $\Delta_{a,x}(k)$ and $\Delta_{a,y}(k)$ are the $k_x$- and $k_y$-type 
solutions on band $a$, respectively. 
The gap function predicted by the third-order perturbation theory 
is presented in Fig.~\ref{Fig3_5}. 
The gap magnitude takes the maximum value on the $\gamma$ band 
and the minimum on the $\beta$ band, reflecting the dominance of the $\gamma$ band. 
The calculated result predicts that the superconducting gap should be strongly anisotropic 
in the basal plane: gap minimum on the $\gamma$ Fermi surface should exist 
near the zone boundary, and nearly-zero gap minima 
on the $\alpha$ and $\beta$ Fermi surfaces should exist at the diagonal points. 
The gap minimum on the $\gamma$ band occurs intrinsically 
near the zone boundary points $(\pm\pi, 0)$ and $(0, \pm\pi)$, 
since $\Delta(k)=0$ holds at the points $(k_x, k_y)=(0, 0)$, $(\pm\pi, 0)$, $(0, \pm\pi)$
and $(\pm\pi, \pm\pi)$ in the Brillouin zone, as straightforwardly verified from odd-parity symmetry 
and $2\pi$-periodicity of $\Delta(k)$~\cite{Nomura2000a, Miyake1999}. 
We should note that the nearly-zero gap minima on the passive $\alpha$ and $\beta$ surfaces 
are not a direct result from the symmetry property, 
in other words, not due to the sign change of the gap function, 
in contrast to the nodes of $d_{x^2-y^2}$-wave pairing. 
The reason why such a strongly anisotropic gap structure appears 
on the passive $\alpha$ and $\beta$ bands 
is that the antiferromagnetic fluctuation due to the nesting 
between the $\alpha$ and $\beta$ Fermi surfaces weakens 
the triplet pairing at these points. 
Overall, the anisotropy of the main gap structure is quite different 
from that of the $d_{x^2-y^2}$ symmetry

\begin{figure}
\begin{center}
\includegraphics[width=3.2in]{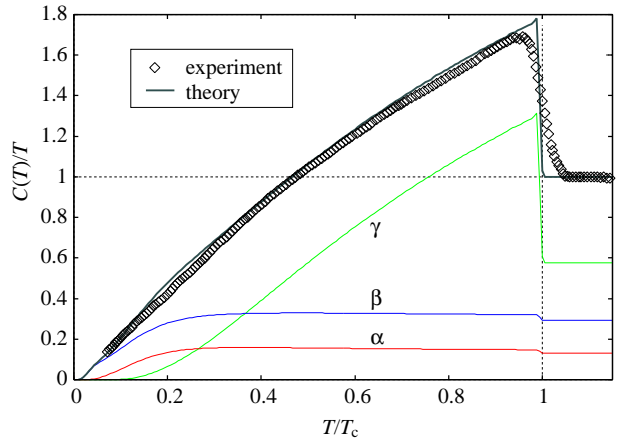}
\end{center}
\caption{(Color online) 
Specific heat $C$ divided by temperature $T$ is displayed 
as a function of temperature (normalized by the normal-state value)~\cite{Nomura2002c}. 
The solid curves and the diamonds ($\diamond$) represent the theoretical 
and experimental data, respectively. 
Each contribution of the three bands is also depicted separately.}
\label{Fig3_6}
\end{figure}

Nomura and Yamada calculated the specific heat 
using the above gap function~\cite{Nomura2002c}. 
They assumed that the momentum dependence 
of superconducting gap near $T_{\rm c}$ is preserved 
down to low temperatures, and simply determine 
the absolute magnitude of the gap 
below $T_{\rm c}$ by the standard BCS gap equation. 
The calculated result is shown in Fig.~\ref{Fig3_6}, 
and is compared 
with experimental results~\cite{Deguchi2004, NishiZaki2000}. 
The jump of $C/T$ at $T_{\rm c}$ is dominated by the contribution 
from the $\gamma$ band. 
Note that, owing to the strongly anisotropic gap structure, 
the $T$-linear behavior at low temperatures is reproduced 
even if the chiral symmetry is assumed. 
At low temperatures, thermal excitations on the $\alpha$ and $\beta$ bands, 
dominantly contribute to the specific heat, due to the nearly-zero gap minima 
on these passive bands. 
The same gap function can explain comprehensively the temperature dependences 
of NMR relaxation rate ($1/T_1T$)~\cite{Nomura2008}, 
ultrasound attenuation rate, and thermal conductivity~\cite{Nomura2005}. 

\subsection{Role of spin-orbit interaction and $d$-vector orientation}
\label{RoleSOI}

Electron-electron Coulomb interaction is symmetric under rotations in spin space. 
Therefore, pairing interaction $V(k, k')$ induced only by the Coulomb repulsion 
does not bring about any anisotropy in spin space, 
and never lifts the degeneracy of the $d$-vector states in principle. 
In \S~\ref{dvecSec}, three major contributions to the anisotropic energy for the $d$-vector have been discussed,
as summarized in Table~\ref{dvec}.
Among them, we give here a detailed discussion on a leading mechanism of lifting the degeneracy, 
the ``atomic'' SOI among the Ru-$4d$ orbitals. 
This interaction hybridizes different spin and orbital states with each other 
and causes anisotropy in spin space~\cite{Ng2000,Yanase2003b,
Annett2006,Yoshioka2009}. 
Although the SOI is believed to be much weak compared 
with the bandwidth and the Coulomb interaction for the Ru-$4d$ states 
(of the order of 100 meV), it plays the essential role 
for determining the most stable $d$-vector state. 
The degenerated $d$-vector states generally split into five states 
due to the SOI, as shown in Table~\ref{table3}. 

Here we review several microscopic theories including SOI. 
In the early stage, Ng and Sigrist~\cite{Ng2000} showed that, 
under reasonable assumptions, 
the SOI among the Ru-$4d\varepsilon$ orbitals 
stabilizes the chiral pairing state, where they used 
$\sin k_x$ and $\sin k_y$ basis functions for the gap function. 
They used a realistic three-band model including nearest-neighbor ferromagnetic exchange. 
According to their results, if the pairing interaction is dominant 
on the $\gamma$ band, then the chiral state ${\mib d}({\mib k}) =
\hat{z}\Delta_0(\sin k_x \pm {\rm i} \sin k_y)$ is reasonably the most stable against the others. 

\begin{figure}
\begin{center}
\includegraphics[width=3in]{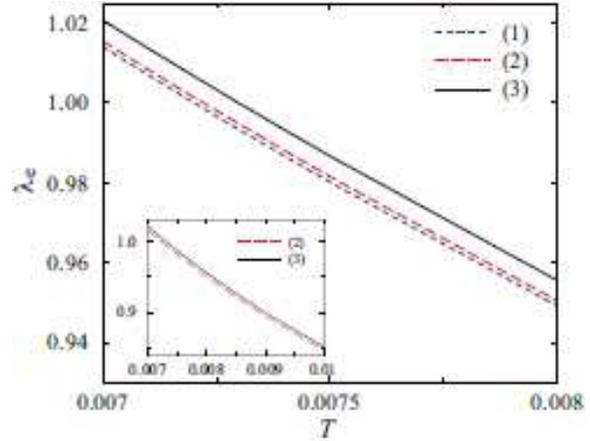}
\end{center}
\caption{(Color online) 
Eigenvalues $\lambda_{\rm e}$ as a function of temperature 
for the five $d$-vector states~\cite{Yanase2003b}: 
(1) ${\mib d}({\mib k}) \propto \hat{x}k_x \pm \hat{y}k_y$, 
(2) ${\mib d}({\mib k}) \propto \hat{x}k_y \pm \hat{y}k_x$,
(3) ${\mib d}({\mib k}) \propto \hat{z}(k_x \pm  {\rm i} k_y)$. 
The inset shows the same in the wide temperature region.}
\label{Fig3_7}
\end{figure}

To investigate the possibility of the chiral pairing state in a more microscopic level, 
Yanase and Ogata included the SOI to perturbation calculation 
for a three-band Hubbard model~\cite{Yanase2003b}. 
They treated the SOI within second-order perturbation. 
For pairing interaction, they performed perturbation expansion 
in the intra-orbital and inter-orbital Coulomb integrals 
to third and second order, respectively. 
The origin of the attractive interaction will be the same as that 
by Nomura and Yamada, i.e., the third-order vertex-corrected terms 
with respect to the intra-orbital Coulomb repulsion $U$. 
The calculated eigenvalues of the linearized Eliashberg equation 
are shown in Fig.~\ref{Fig3_7}, where note that the $d$-vector state 
presenting the larger eigenvalue is the more favorable state. 
As seen in Fig.~\ref{Fig3_7}, the chiral pairing state 
${\mib d}({\mib k}) \propto \hat{z}(k_x \pm {\rm i} k_y)$ is expected to be realized 
at $T_{\rm c}$. 
This is an ESP state with the pair spins in the basal plane.  

One might naively consider that the magnetic field necessary to change the direction of the $d$-vector is estimated from the SOI energy $\lambda$
via a simple relation $\mu_\mathrm{B}B \sim \lambda$.
Such estimation is not correct because the anisotropy energy should be evaluated from 
the difference in condensation energy among states with different $d$-vector orientations.
This energy should be of the order of $k_\mathrm{B}\Delta\Tc$, 
where the difference in transition temperature $\Delta \Tc$ turns out to be a function of $\lambda/W$ rather than $\lambda$ itself, 
as mentioned below.

Yanase and Ogata showed that realistic value of spin-orbit coupling 
(about 50 meV) lifts the degeneracy only slightly.
This is because the first-order contribution with respect to the spin-orbit coupling 
vanishes and the symmetry-breaking energy is of the order of $UJ\lambda^2/W^3$ 
($U$: intra-orbital Coulomb repulsion, $J$: Hund's rule coupling, and $W$: bandwidth), 
if the hybridization between different orbitals is completely neglected. 
According to their estimation, the splitting of $T_{\rm c}$ 
between different $d$-vector states is estimated 
as $\Delta T_{\rm c} \sim  0.04 T_{\rm c}$. 
A very crude estimation using $\mu_\mathrm{B}B \sim k_\mathrm{B}\Tc \cdot (\Delta \Tc/\Tc)$ leads to 90~mT.
Such slight lifting of degeneracy means that the $d$-vector 
is not so strongly pinned along the most stable direction 
and can be rotated easily even by a small magnetic field. 
In fact, the $d$-vector direction must be rotated 
even by a small magnetic field, to be qualitatively consistent 
with the experimental fact that the NMR Knight shift 
remains unchanged at $T_{\rm c}$ under a small magnetic field 
(20~mT) along the $c$ axis~\cite{Murakawa2007}. 

\begin{figure}
\begin{center}
\includegraphics[width=3.2in]{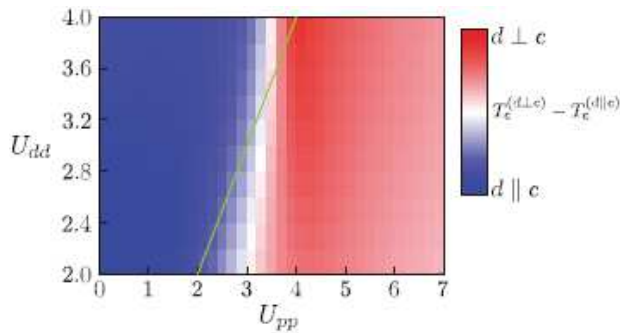}
\end{center}
\caption{(Color online) 
Phase diagram of anisotropy of the $d$-vector~\cite{Yoshioka2009}. 
The stable direction of the $d$-vector is parallel (perpendicular) 
to the $c$ axis in the blue (red) region. 
The green solid line denotes $U_{dd}=U_{pp}$ points.}
\label{Fig3_8}
\end{figure}

More recently, Yoshioka and Miyake investigated 
the anisotropy of the $d$-vector on the basis of the $d$-$p$ model, 
including not only the on-site Coulomb repulsion $U_{dd}$ 
and SOI at each Ru site 
but the on-site Coulomb repulsion $U_{pp}$ at each O site~\cite{Yoshioka2009}. 
They calculated the pairing interaction within the perturbation theory 
with respect to the on-site Coulomb repulsions $U_{dd}$ and $U_{pp}$ 
and the SOI $\lambda$. 
They found that, if $U_{pp}$ is assumed to be larger than $U_{dd}$, 
the $d$-vector parallel to the $ab$ plane can be more stable 
than the one parallel to the  $c$ axis, as shown 
in the phase diagram of Fig.~\ref{Fig3_8}. 
According to their estimation, the splitting of $T_{\rm c}$ 
between the ${\mib d} \parallel c$ and ${\mib d} \perp c$ states 
due to the SOI is $\Delta T_{\rm c} \sim 0.022T_{\rm c}$ 
for $\lambda=0.2$ and $U_{dd}=U_{pp}=4.0$ 
(in units of $t_{dp}$, where $t_{dp}$ is the hopping 
between nearest-neighbor Ru and O sites). 
In addition, they showed that the momentum dependence of the order parameter 
changes from $\sin k_x\cos k_y$-like to simple $\sin k_x$-like, 
around $U_{pp} \sim 3 t_{dp}$, as $U_{pp}$ is increased. 
However, we should note that, according to Table~\ref{table3}, 
any of $d$-vector states with ${\mib d} \perp c$ 
do not break TRS in the absence of magnetic fields, 
and therefore seem incompatible with $\muup$SR results~\cite{Luke1998}. 

At present, the mechanism based on the Coulomb repulsion 
at Ru site (the main band is $\gamma$), 
incorporated with the atomic SOI at Ru site, 
reasonably supports the realization of the chiral pairing state. 
To be consistent with the NMR Knight shift measurement~\cite{Murakawa2007}, 
it is necessary to assume that the $d$-vector along the $c$ axis 
should be rotated to the $ab$ plane even under a small magnetic field 
of as small as 20~mT and the Coulomb interaction at oxygen site 
should not be so large as the $d$-vector is laid along the $ab$ plane. 
Of course, the microscopic determination of the most stable $d$-vector state 
remains still a subtle issue, because the stabilization energy of the direction 
of the $d$-vector is quite small and moreover possibly competing, as mentioned above. 

We would like to comment on one of the promising future directions of the microscopic mechanism theory.
In all the theories discussed here, simple two-dimensional tight-binding electronic structures are adopted.
To proceed to a more precise evaluation of the mechanism and $d$-vector states,
one should use more precise electronic structures based on first-principle calculations.
Extension to three-dimensional electronic structures may also prove crucial to resolve some of the important issues,
such as suppression of $\Hcc$ and superconducting multiphase, discussed in \S~\ref{unresolved}.

Recent LDA calculations of the electronic structure and ARPES measurements \cite{Haverkort2008PRL, Iwasawa2010PRL}
suggest that the three cylindrical Fermi surfaces cross near the $\langle 110 \rangle$ plane 
in the absence of the SOI, 
while actually the crossing is removed by the SOI. 
According to these results, the orbital character on the Fermi surface varies with $k_z$, 
with a tendency of crossing between two-dimensional and one-dimensional bands near $k_z=0$.
Thus we can no longer consider that the $\gamma$ Fermi surface is constructed almost entirely from the $xy$ electrons. 
Influences of such Fermi surface crossing have not been studied so far, 
and remain an interesting issue to be investigated in detail. 


\section{Challenges to the spin-triplet scenario}
\label{unresolved}
In this section, we discuss some of the unresolved issues for which explanations 
in terms of the present spin-triplet $p$-wave scenario have not been settled. 
We discuss available theoretical models. 

\subsection{Suppression of the upper critical field}
\label{Hc2limit}
The $\Hcc$ of \sro\ is strongly anisotropic, reflecting the quasi-two-dimensional feature of the Fermi surface: 
1.5~T for $H \parallel ab$ and 0.075~T for $H \parallel c$. 
The in-plane $\Hcc$ anisotropy appears below around 1~K and increases up to 0.04~T. 
One of the unresolved issues on \sro\ is the temperature dependence of $\Hcc$ for the field in the $ab$ plane. 
If the orbital depairing effect dominates in determining $\Hcc$, 
$\Hcc$ is expected to increase linearly with the initial slope at $\Tc$ on cooling and 
saturates to $-0.7\Tc\mathrm{d}\Hcc/\mathrm{d}T|_{T=\Tc}$ at low temperatures. 
This behavior is described by the so-called Werthamer-Helfand-Hohenberg (WHH) theory \cite{Helfand1966PR,Werthamer1966PR} and 
its extension to $p$-wave superconductors \cite{Maki1999JS,Lebed2000PhysicaC}. 
Although $\Hcc$ of \sro\ for $H \parallel c$ exhibits conventional behavior ($\Hcc(0)=-0.78\Tc\mathrm{d}\Hcc/\mathrm{d}T|_{T=\Tc}$), 
the in-plane $\Hcc$ is strongly suppressed at low temperatures ($\Hcc(0)=-0.42\Tc\mathrm{d}\Hcc/\mathrm{d}T|_{T=\Tc}$) \cite{Kittaka2009JPCS-2}. 

\begin{figure}
\begin{center}
\includegraphics[width=3.2in]{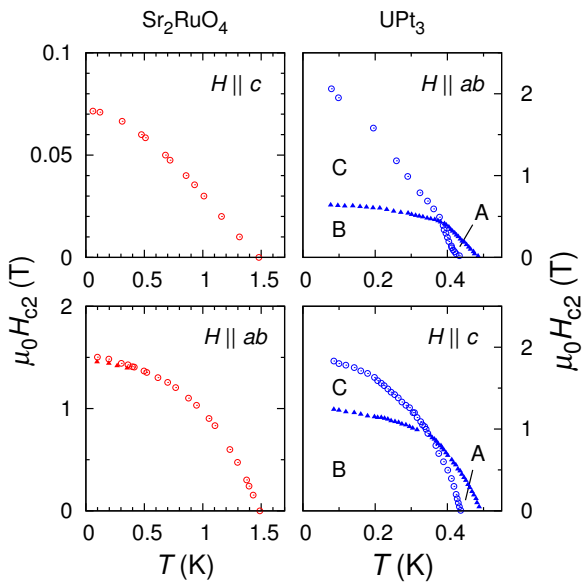}
\caption{
(Color online) Field-temperature phase diagrams of \sro\ compared with those of UPt$_3$. 
Suppression of $\Hcc$ of similar magnitude is evident for $H \parallel ab$ in \sro\ and for $H \parallel c$ in UPt$_3$. 
From Deguchi \textit{et al}. (2002, 2004) \cite{Deguchi2002JPSJ,Deguchi2004} and Dijk \textit{et al}. (1993).\cite{Dijk1993JLTP}
}
\label{Fig4-1}
\end{center}
\end{figure}

Recently, the Pauli paramagnetic effect was suggested as a possible origin to limit the in-plane $\Hcc$ of \sro \cite{Machida2008PRB,Choi2010JKPS}. 
Machida and Ichioka calculated $\Hcc(\theta)$ as well as the field dependence of the specific heat and magnetization 
using the Eilenberger equation based on a spherical Fermi surface with the assumption of the paramagnetic depairing effect. 
Although they suggested that their results reproduced some experimental results, 
the Pauli effect occurs only for either the singlet pairing state or triplet pairing states with the $d$-vector locked in a particular direction in the basal plane, 
which is inconsistent with the results of Knight shift experiments suggesting the triplet pairing with the $d$-vector state 
pointing perpendicular to the field in any field direction. 
The NMR result strongly indicates that Pauli effect is not operative in \sro. 
Qualitatively similar $\Hcc$ limiting is also known for another spin-triplet superconductor UPt$_3$ for $H \parallel c$ (Fig.~\ref{Fig4-1}) \cite{Joynt2002RMP,Machida1998JPSJ}, 
whose origin has not been resolved either. 
In the field direction in which a strong $\Hcc$ suppression has been observed,
$H \parallel ab$ for \sro\ and $H \parallel c$ for UPt$_3$, 
the $d$-vector is believed to be perpendicular to the field and thus the Pauli paramagnetic depairing is not anticipated.
Thus, the origins of these $\Hcc$ limitings observed in two plausible spin-triplet superconductors remain challenging issues to be resolved. 

\begin{figure}
\begin{center}
\includegraphics[width=3.2in]{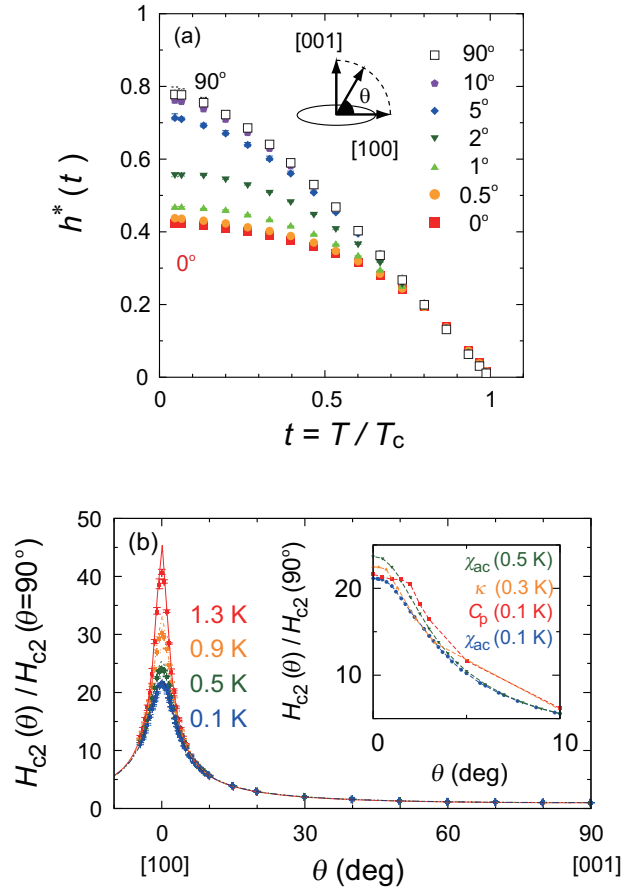}
\caption{
(Color online) (a) Anisotropy of the upper critical fields $\Hcc$ of \sro. 
$h^\ast$ is $\Hcc$ normalized by the initial slope in the vicinity of $\Tc$.
Strong suppression of $\Hcc$ occurs for the polar angle only within a few degrees from the basal plane. 
(b) Field-angle $\theta$ dependence of $\Hcc$ normalized by $\Hcc(\theta=90^\circ)$. 
The $\Hcc$ suppression at $\theta=0^\circ$ is evident even at relatively high temperatures.
From Kittaka \textit{et al}. (2009). \cite{Kittaka2009PRB}
The inset compares the $\Hcc(\theta)$ determined from various experiments.
}
\label{4-1_H-T_phase}
\end{center}
\end{figure}

In order to clarify the detailed features of the $\Hcc$ limiting in \sro, 
Kittaka \textit{et al}. performed ac susceptibility measurements with a precise control of the magnetic field direction  
using a vector magnet system \cite{Kittaka2009PRB}. 
They investigated the dependence of $\Hcc$ on temperature and the angle $\theta$ between the magnetic-field direction and the $ab$ plane, 
as shown in Figs.~\ref{4-1_H-T_phase}(a) and \ref{4-1_H-T_phase}(b). 
It was revealed that the $\Hcc$ limiting is remarkable only for $|\theta| < 5^\circ$ and 
remains strong even at relatively high temperatures close to $\Tc$. 
In the previous study, a clear kink in $\Hcc(\theta)$ around $\theta=2^\circ$ was reported 
from the specific heat measurements at 0.1~K \cite{Deguchi2002JPSJ}, 
but such a clear feature was not detected at any temperature either by the thermal conductivity \cite{Deguchi2002JPSJ} 
or recent ac susceptibility \cite{Yaguchi2002PRB, Kittaka2009PRB} (the inset of Fig.~\ref{4-1_H-T_phase}(b)). 
It was also revealed that the angular dependence of $\Hcc$ determined from the ac susceptibility was reproduced 
with an effective-mass model for an anisotropic three-dimensional superconductor, 
though the anisotropic ratio needs to vary with temperature \cite{Kittaka2009PRB}. 
It is essential to develop a new theory which explains the observed $\Hcc$ behavior as well as other key experimental results, 
such as invariance of the Knight shift.

\subsection{Superconducting multiphase}
\begin{figure}
\begin{center}
\includegraphics[width=2.6in]{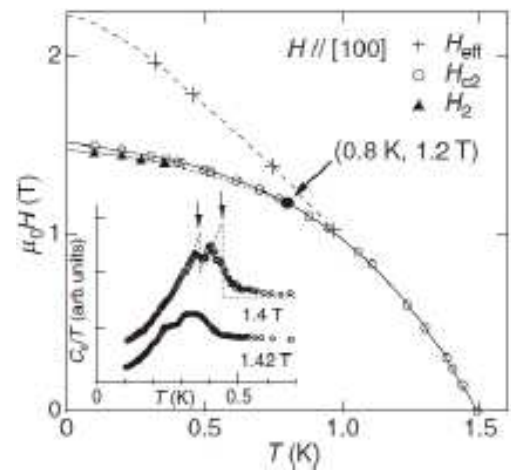}
\caption{
Field-temperature phase diagram of \sro\ detailing 
the regions of superconducting double transitions for field accurately aligned within the basal plane. 
The phase diagram is constructed from the double-peak structure of the temperature dependence of the specific heat. 
From Deguchi \textit{et al}. (2002). \cite{Deguchi2002JPSJ}
}
\label{4-2_multiphase}
\end{center}
\end{figure}

A multi-component superconducting order parameter leads to an additional phase transition in magnetic fields. 
For example, another spin-triplet superconductor UPt$_3$ is well established to have additional superconducting transition in magnetic fields, 
resulting from its multi-component order parameter. 
Also in \sro, an additional transition was observed just below $\Hcc$ at low temperatures below 0.8~K (Fig.~\ref{4-2_multiphase}) 
by the specific heat \cite{Deguchi2002JPSJ}, dc magnetization \cite{Tenya2006JPSJ}, and ac susceptibility~\cite{Mao2000PRL,Yaguchi2002PRB}. 
Intriguingly, this additional transition as well as the in-plane $\Hcc$ anisotropy is suppressed 
by a slight misalignment of the field from the $ab$ plane; 
they are observed only for $\theta \lesssim 1^\circ$.\cite{Mao2000PRL}

Although the mechanism of this superconducting multiphase has been actively discussed, it is still controversial. 
For a superconductor with $\Vec{d}(\mib{k}) \propto \hat{z}(k_x \pm {\rm i} k_y)$, 
if the magnetic field is applied parallel to the $x$ axis, 
the degeneracy between $\hat{z}k_x$ and $\hat{z}k_y$ is lifted and the transition 
from $\hat{z}(k_x \pm {\rm i} \varepsilon k_y)$ to $\hat{z}k_x$ occurs 
and the non-chiral single-component $\hat{z}k_x$ is expected to become stable near $\Hcc$ at all temperatures \cite{Kaur2005PRB}. 
This additional transition can be easily suppressed by a slight misalignment of the field of only one degree. 
However, in this theoretical model, large in-plane anisotropy for such additional transition is expected \cite{Kaur2005PRB,Mineev2008PRB}, 
which is incompatible with the experimental results. 
Another possibility is the crossover from $\hat{z}k_x$ to the spin-polarized $(\hat{z}- {\rm i} \hat{y})k_x$ state \cite{Udagawa2005JPSJ}. 
In this theory, the transition from $\hat{z}(k_x \pm {\rm i} k_y)$ to $\hat{z}k_x$ is additionally expected in the region far below $\Hcc$. 
Although a tiny magnetization anomaly far below $\Hcc$ was reported by Tenya \textit{et al}. \cite{Tenya2006JPSJ}, 
there is no other report which supports the results of the dc magnetization experiment. 
At present, this scenario is not well established, either. 
Machida and Ichioka proposed that the second transition is attributed to the FFLO in Pauli limited superconductors \cite{Machida2008PRB}. 
Clearly, there remains a contradiction between the Pauli effect scenario and the results of Knight shift experiments, 
as written in \S~\ref{Hc2limit}.

\section{Novel superconducting phenomena}
\subsection{Superconductivity in eutectic systems}

The intrinsic $\Tc$ of pure \sro\ is well established to be 1.5~K \cite{Mackenzie1998PRL}. 
However, a large diamagnetic signal and the rapid decrease of resistivity with an onset $\Tc$ of nearly 3~K are observed 
in some single crystals of \sro. 
Soon after the discovery of such ''3-K phase'', 
it was noticed that the enhanced-$\Tc$ superconductivity originates in the \sro-Ru eutectic solidification system \cite{Maeno1998PRL}. 

\begin{figure}
\begin{center}
\includegraphics[width=2.3in]{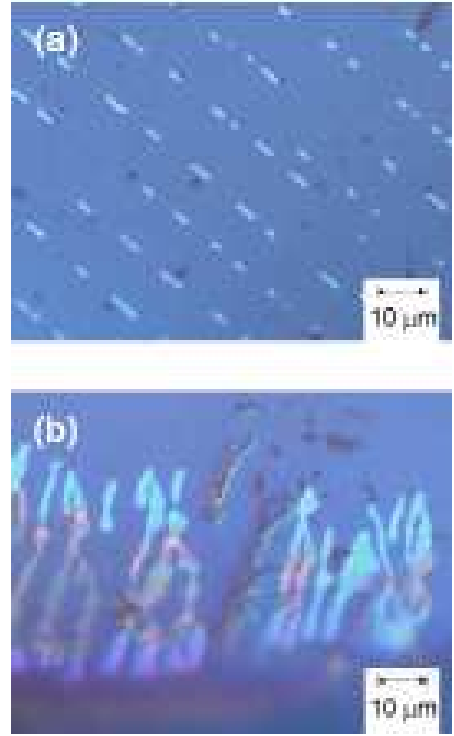}
\caption{
(Color online) Optical microscope pictures of the surface of a \sro-Ru eutectic crystal. Thin Ru metal lamella are embedded in the matrices of \sro\ single crystal. 
From Maeno \textit{et al}. (1998). \cite{Maeno1998PRL}
}
\label{5-1_3-K_PLOM}
\end{center}
\end{figure}

\begin{figure}
\begin{center}
\includegraphics[width=3in]{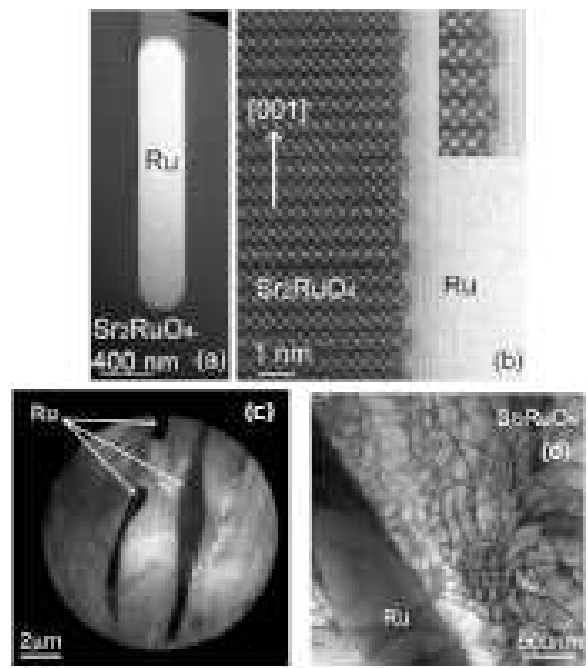}
\caption{
Transmission electron microscopy images of \sro-Ru eutectic crystals.
From Ying \textit{et al}. (2009). \cite{Ying2009PRL}
Note that the dislocation lines in (d) are the results of superposition over many layers.
}
\label{5-1_3-K_TEM}
\end{center}
\end{figure}

The \sro-Ru eutectic system is obtained during the single crystal growth of \sro\ with a floating zone technique using Ru self-flux \cite{Mao2000MRB}. 
Because volatile RuO$_2$ evaporates from the surface of melt during the growth, 
excess Ru needed as a flux tends to solidify in the core region of the crystal rod and coexists with \sro,
forming a periodic pattern characteristic of eutectic solidification. 
Figures~\ref{5-1_3-K_PLOM}(a) and \ref{5-1_3-K_PLOM}(b) represent polarized optical microscope images of polished surfaces of a eutectic crystal 
parallel and perpendicular to Ru lamellae, respectively \cite{Maeno1998PRL}. 
A number of Ru lamellae (bright area) with typical dimensions of $30 \times 10 \times 1$~$\muup$m$^3$ are embedded 
in a \sro\ single crystal (dark area). 
The distance between the nearest lamellae is on typically about 10~$\muup$m. 
The orientation of Ru lamellae with respect to the crystal axes of \sro\ varies among different crystals and even within the same crystalline rod. 
Surprisingly, although the values of the unit-cell parameters are different between \sro\ and Ru, 
it was revealed from transmission electron microscopy (TEM) images that the interface between Ru and \sro\ is atomically sharp, 
as shown in Figs.~\ref{5-1_3-K_TEM}(a) and \ref{5-1_3-K_TEM}(b) \cite{Ying2009PRL}. 
By contrast, dislocations caused by the lattice mismatch are distributed on a larger scale, 
as displayed in TEM pictures as dark lines (Figs.~\ref{5-1_3-K_TEM}(c) and \ref{5-1_3-K_TEM}(d)). 

Various experimental results suggest that enhanced-$\Tc$ superconductivity is non-bulk and non-$s$-wave, 
occurring in the \sro\ region in the eutectic system. 
First, a broad transition with a large diamagnetic shielding \cite{Maeno1998PRL} and a tiny anomaly of the specific heat around 3~K \cite{Yaguchi2003PRB} indicate that 
the 3-K superconductivity is non-bulk. 
Second, a zero-bias conductance peak (ZBCP) observed above the bulk $\Tc$ of \sro\ in the \sro-Ru junctions suggests that
non-$s$-wave superconductivity occurs in the 3-K phase \cite{Mao2001PRL, Kawamura2005JPSJ, Yaguchi2006JPSJ}. 
Third, the anisotropy of $\Hcc$ clearly depends on the orientation of the crystal axes of \sro\ rather than that of Ru lamellae \cite{Maeno1998PRL,Ando1999JPSJ}, 
indicating that \sro\ is a stage of enhanced-$\Tc$ superconductivity. 
In the 3-K phase, the formation of Josephson network consisting of inter-lamellar supercurrents is suggested by 
the zero-voltage current observed at temperatures well above the bulk $\Tc$ of \sro\ \cite{Hooper2004PRB} and
a large diamagnetic shielding with a characteristic response to ac and dc magnetic fields \cite{Kittaka2009JPSJ}. 

One of the peculiar features of the 3-K phase is temperature dependence of $\Hcc$ \cite{Wada2002JPCS}. 
Using a definition of $\Hcc$ as the resistivity inflection point associated with the superconducting transition to the 3-K phase, 
$\Hcc$ appears to behave as $(1-T/\Tc)^{0.72}$ for $H \parallel c$ and $(1-T/\Tc)^{0.75}$ for $H \parallel ab$ close to $\Tc$ \cite{Yaguchi2003PRB}. 
In addition, $\Hcc(T)$ for $H \parallel c$ was revealed to have an upward curvature around 2~K. 
Such unusual behavior of $\Hcc$ is explained using a phenomenological theory within the Ginzburg-Landau formalism 
for an interfacial superconductivity by Sigrist and Monien (S-M) \cite{Sigrist2001JPSJ}. 
They assumed one interface and analyzed with the vector order parameter $\Vec{d}(\mib{k})=\hat{z}(\eta_x k_x + \eta_y k_y)$, 
where $(\eta_x,\eta_y)=(1,{\rm i})$ is plausible in \sro. 
In the S-M model, it was suggested that 
the presence of interface parallel to the $y$ axis makes one of the order parameters $k_y$ more stable than the other $k_x$. 
By introducing a finite width of the interfacial superconductivity in the S-M model, 
Matsumoto \textit{et al}. succeeded in reproducing the unusual temperature dependence of $\Hcc$ more quantitatively \cite{Matsumoto2003JPSJ}. 
However, the mechanism of the strong in-plane $\Hcc$ suppression at low temperatures and 
the origin of the hysteresis behavior observed just below $\Hcc$ for $H \parallel ab$ remain unresolved. \cite{Ando1999JPSJ}

By analyzing the interplay between the chiral $p$-wave order parameter of \sro\ and 
the $s$-wave order parameter of a cylindrical Ru inclusion based on the Ginzburg-Landau formulation, 
Kaneyasu and Sigrist pointed out that, at a temperature somewhat lower than $\Tc$ of Ru ($\sim 0.49$~K), 
a state emerges with a spontaneous flux distribution around a Ru inclusion \cite{Kaneyasu2010JPSJ}. 
Moreover, Kaneyasu \textit{et al}. proposed that 
a discontinuous transition from time-reversal-symmetry-conserving phase to time-reversal-symmetry-breaking phase occurs on cooling 
at a temperature between 1.5 and 3~K \cite{Kaneyasu2010JPSJ-2}. 
Their model explains anomalous asymmetric features with respect to the current direction in the current-voltage characteristics observed below 2.3~K \cite{Hooper2004PRB} 
as well as the observation of ZBCP below 2.4~K \cite{Kawamura2005JPSJ}. 

\begin{figure}
\begin{center}
\includegraphics[width=3.2in]{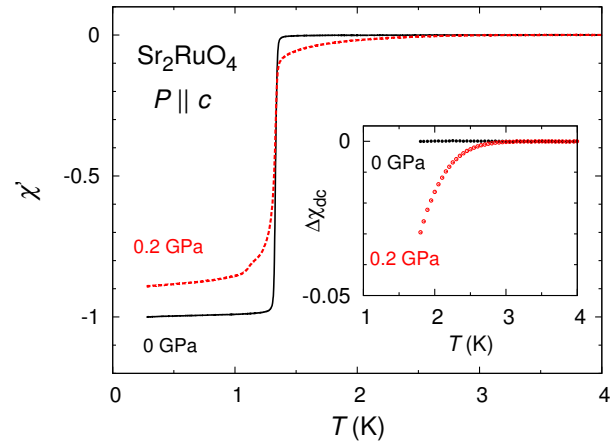}
\caption{
(Color online) Diamagnetic shielding signals of a pure \sro\ crystal under uniaxial pressure. 
Evidence for the enhancement of superconducting transition temperature up to 3~K is clearly visible. 
From Kittaka \textit{et al}. (2010). \cite{Kittaka2010PRB}
}
\label{5-1_214UP}
\end{center}
\end{figure}

Recently, hints for clarifying the mechanism of the enhancement of $\Tc$ were provided by Kittaka \textit{et al}. 
from uniaxial pressure experiments \cite{Kittaka2010PRB}. 
They found that uniaxial pressure along the $c$ axis of only 0.2~GPa induces 
a broad superconducting transition with the onset $\Tc$ of about 3.2~K 
even in a pure, Ru-inclusion-free \sro\ sample with a sharp transition at $\Tc = 1.35$~K at ambient pressure. 
The results of the ac susceptibility measurements are shown in Fig.~\ref{5-1_214UP}. 
It was also found that both out-of-plane and in-plane uniaxial pressures largely enhance the magnetic shielding fraction associated with 3-K superconductivity 
in eutectic samples \cite{Yaguchi2009JPCS,Kittaka2009JPSJ-2}. 
In particular, in-plane uniaxial pressure is much more effective to the enhancement. 
By contrast, hydrostatic pressure suppresses the bulk $\Tc$ of \sro\ \cite{Shirakawa1997PRB,Forsythe2002PRL} 
as well as the shielding fraction associated with 3-K superconductivity \cite{Yaguchi2010JPSJ}. 
These facts strongly indicate that an anisotropic crystal distortion of \sro\ enhances its $\Tc$. 
Nevertheless, the microscopic origin of this enhancement 
by anisotropic distortion has not been clarified.

Let us describe another \sro-based eutectic system, \sro-\SRO. 
Fittipaldi \textit{et al}. found that \sro-\SRO\ eutectic crystals can be grown with a feed rod containing 45\% excess Ru 
with respect to \sro\ under a gas mixture of 10\% oxygen and 90\% argon with the total pressure of 10 atm. \cite{Fittipaldi2005JCG}
X-ray diffraction analyses of \sro-\SRO\ eutectic crystals indicate that the directions not only of the $c$ axis 
but also of the in-plane axes of \sro\ and \SRO\ are common in the eutectic system \cite{Fittipaldi2005JCG}. 
It was also revealed from TEM images that interfaces parallel to the growth direction 
(longitudinal direction of a crystalline rod) are sharp and defect-free 
whereas interfaces perpendicular to the growth direction appear wavy and decorated with Ru precipitates of micron-size \cite{Ciancio2009APL}. 
In addition, mono-layers of RuO$_2$ planes with a thickness on the order of nm (very thin \sro\ layer) were revealed 
to be embedded as stacking faults in the apparent \SRO\ regions of the eutectic system. 
Nevertheless, the \SRO\ domains of the eutectic system have a significantly lower level of impurities, 
such as Sr$_4$Ru$_3$O$_{10}$ ($n=4$) and SrRuO$_3$ ($n=\infty$), compared to \SRO\ single-phase crystals \cite{Ciancio2009PRB}. 

In ac susceptibility measurements with an apparent \SRO\ domain cut from the \sro-\SRO\ eutectic system, 
multiple superconducting transitions at 1.3 and 1~K were observed \cite{Kittaka2008PRB, Kittaka2009JPCS-1}. 
These transitions were revealed to be non-bulk from specific heat measurements and are considered to be attributed to 
the superconducting transitions of thin \sro\ layers embedded in the \SRO\ domain as stacking faults. 
In fact, the anisotropy of $\Hcc$ of these transitions is the same as that of the bulk \sro\ and
$\Hcc$ for $H \parallel c$ of the transition at 1.3~K (1~K) is compatible with (slightly lower than) that of the bulk \sro\ \cite{Kittaka2008PRB, Fittipaldi2008EPL}.

At the early stage, the possibility of the long-range proximity effect in the eutectic system 
arising from the bulk \sro\ region into \SRO\ was discussed 
because finite critical current was observed in a crystal containing a large amount of \SRO \cite{Hooper2006PRB}. 
This finite critical current is probably not due to the long-range proximity effect, 
but due to the presence of very thin \sro-layer inclusions. 
In fact, $\Tc$ of the multiple superconducting transitions is not very sensitive to the electronic state of \SRO;
although \SRO\ exhibits ferromagnetism under the uniaxial pressure along the $c$ axis above 0.3 GPa \cite{Ikeda2004JPSJ,Yaguchi2006AIP}, 
the onset temperature of the superconducting transition at 1~K is almost invariant or increases slightly under the uniaxial pressure up to 0.4~GPa \cite{Kittaka2010PhysicaC}.
This uniaxial pressure effect is consistent with the scenario that the multiple transitions originate from the embedded thin \sro\ layers.

Recently, Yanase focused on a model of a thin \sro\ layer intercalated in the \SRO\ domain 
as a multi-component spin-triplet superconductor with the random spin-orbit coupling. \cite{Yanase2010JPSJ}
He proposed that the $d$-vector state in such a \sro\ layer is different from that in the bulk \sro\ and 
suggested that the $d$-vector state is locked to be $\Vec{d}(\mib{k}) \propto \hat{x}k_y - \hat{y}k_x$ 
with time-reversal symmetry. 

\subsection{\sro\ as a topological superconductor}

Recently, quantum phenomena for which underlying topology of wave functions plays fundamental roles are actively investigated with a universal point of view. 
Topology is a concept and method enabling the classification of shape by continuous deformation. 
For quantum phenomena, the shape subject to continuous deformation is based on the symmetries and phases of the wave functions. 
A well-known example of topological quantum phenomena is vortex quantization in a superconductor, 
for which phase winding associated with real-space topology determines the state. 
In addition, novel quantum phenomena may emerge in materials for which bulk wave functions themselves have nontrivial topology, 
characterized by winding numbers defined in momentum space; 
thus these materials are called the "topological materials". 
Topological materials include some insulators, superconductors and superfluids and are classified 
in terms of discreet symmetries of the Hamiltonian as well as the special dimensions \cite{Schnyder2008PRB}. 
It has been recognized that states specific to non-trivial topology and immune to disorder emerge at the edge, 
i.e., its surface or interface with other materials, according to ``bulk-edge'' correspondence.\cite{Tanaka2011arX} 

\begin{figure}
\begin{center}
\includegraphics[width=2.4in]{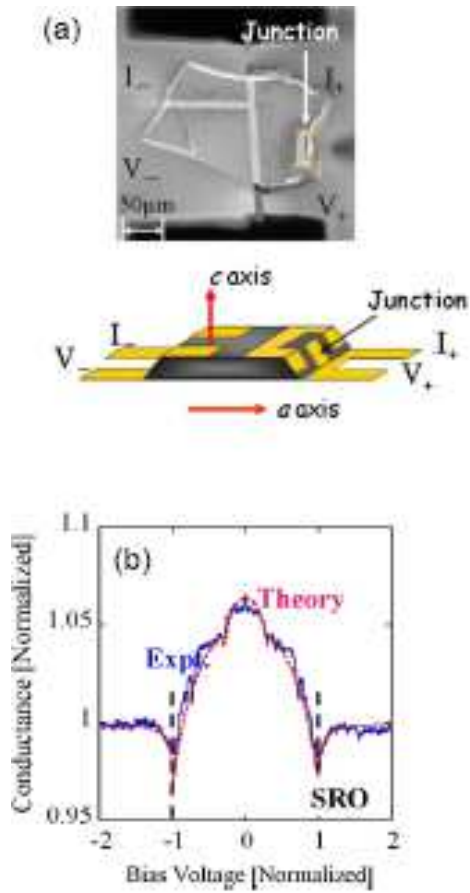}
\caption{
(Color online) 
(a) Scanning ion microscopy image  of a \sro-Au junction fabricated on a SiO$_2$ substrate and a schematic of the junction structure. 
(b) Comparison between experimental conductance spectra at 0.55~K (solid lines) and theoretical spectra (dotted lines). 
From Kashiwaya \textit{et al}. (2011). \cite{Kashiwaya2011PRL}
}
\label{5-2_SINjunction}
\end{center}
\end{figure}

\begin{figure}
\begin{center}
\includegraphics[width=3.2in]{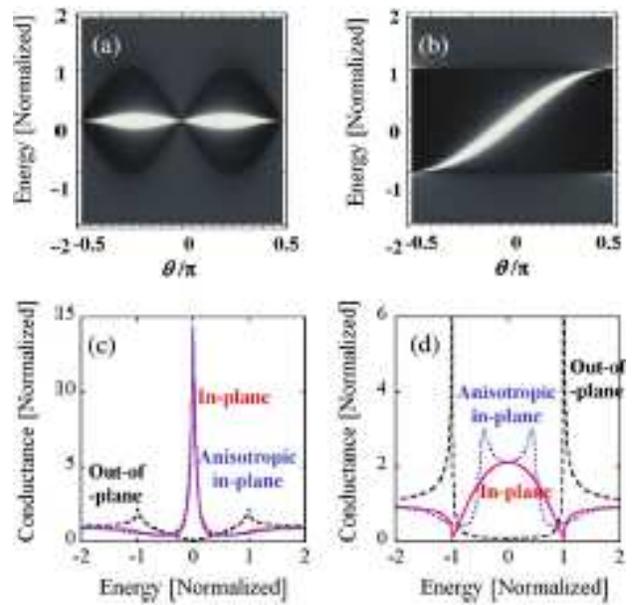}
\caption{
(Color online) 
Angle-resolved conductance spectra for (a) $d_{xy}$-wave superconductors and (b) chiral $p$-wave superconductors. 
The vertical axis is the quasiparticle energy normalized by $\Delta_0$. 
The brightness represents the magnitude of the junction conductance. 
Bright regions correspond to the conductance peak originating from Andreev bound states formed at the edge. 
Conductance spectra for (c) $d_{xy}$-wave superconductors and (d) chiral $p$-wave superconductors. 
From Kashiwaya \textit{et al}. (2011). \cite{Kashiwaya2011PRL}
}
\label{5-2_SINjunction2}
\end{center}
\end{figure}

A chiral $p$-wave superconductor with the wave function $k_x + {\rm i}k_y$ is classified among such topological materials and 
is called a topological superconductor, along with other classes of topological superconductors.
For unitary-spin, chiral $k_x + {\rm i}k_y$ state with broken time-reversal symmetry, 
the topological invariance is specified by the Chern number. \cite{Furusaki2001PRB}
The chiral edge state discussed in \S~\ref{Sec_chiral} is a consequence of bulk-edge correspondence reflecting the bulk topological properties of a chiral $p$-wave superconductor. 
The existence of such chiral edge sate has been demonstrated by a recent experiment of quasi-particle tunneling spectroscopy. \cite{Kashiwaya2011PRL} 
In superconductor-insulator-normal metal junctions with \sro\ and Au as shown in Fig.~\ref{5-2_SINjunction}, broad humps filling up the gap are observed in the conductance spectra. 
Compared with the narrow ZBCP associated with the flat-band edge state of YBa$_2$Cu$_3$O$_7$ at its (110) interface, 
the broad conductance peak is a direct manifestation of the chiral edge state characteristic of chiral $p$-wave superconductivity as shown in Fig.~\ref{5-2_SINjunction2}.
In addition to a number of evidence indicating chiral p-wave symmetry, 
this confirmation of the chiral edge states reassures that Sr$_2$RuO$_4$ is classified as a topological superconductor. 
Very recently, Sasaki \textit{et al}. found that a bulk superconductor Cu$_x$Bi$_2$Se$_3$ \cite{Hor2010PRL,Kriener2011PRL}, 
derived from a doped topological insulator Bi$_2$Se$_3$, exhibits a similar ZBCP in a "soft" point-contact junction 
with gold wire attached on the cleaved surface of a Cu$_x$Bi$_2$Se$_3$ crystal by silver paste \cite{Sasaki2011PRL}. 
After examination of the tunneling spectra in terms of possible superconducting states based on the relatively simple band structure 
expected for Cu$_x$Bi$_2$Se$_3$, they concluded that the observed ZBCP gives crucial evidence 
for topological superconductivity with odd parity in Cu$_x$Bi$_2$Se$_3$.

Such ZBCP emerging at a junction between an unconventional superconductor and a normal metal is a consequence of the density of states due to Andreev bound state. \cite{Tanaka1995PRL}
It has been recognized that such Andreev bound state and the penetration of the pairing amplitude into the normal metal by the proximity effect are intimately related. 
For an unconventional spin-singlet superconductor, 
the formation of the Andreev bound state at the junction surface of the unconventional superconductor and the proximity effect are competitive, 
but it turns out that for a spin-triplet superconductor, they may be constructive, making the proximity effect unusually strong and long-ranged. \cite{Tanaka2004PRL}
As one consequence of such enhancement, it is predicted that due to the proximity penetration of the Andreev bound state into the normal metal, 
in a T-shaped junction consisting of a spin-triplet superconductor and a normal metal, 
the density of state of the normal metal increases below $\Tc$ of the superconductor. \cite{Asano2007PRL}

Moreover, it has been recognized that such Andreev bound state can be described in terms of odd-frequency pairing states. 
In a diffusive normal metal, states other than isotropic $s$-wave states become difficult to survive.
At a junction with a spin-triplet superconductor, therefore, spin-triplet $s$-wave, namely an odd frequency pairing state, is realized. \cite{Tanaka2007PRL-2} 
Intimate relations among surface bound states, proximity effects, odd-frequency pairing states, and topological edge states are discussed in 
a recent review by Tanaka, Sato and Nagaosa. \cite{Tanaka2011arX}

Concerning the superconducting junctions, it has been recognized that small-size crystal is important to extract the intrinsic unconventional behavior of \sro. 
For a focused ion beam (FIB) fabricated small crystal of \sro/Ru eutectic, an unusual hysteresis in the current-voltage characteristics has been observed. 
It was interpreted as due to the motion of the chiral domain walls which alters the critical current. \cite{Kambara2008PRL,Kambara2010JPSJ,Kashiwaya2010PhysicaC}

\begin{figure}
\begin{center}
\includegraphics[width=2.8in]{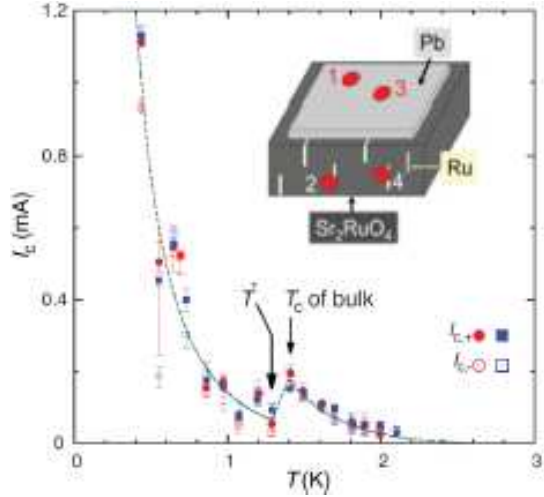}
\caption{
(Color online) 
Variation of the critical current $\Ic$ with temperature of a Pb/Ru/\sro\ junction. 
The inset is a schematic of the junction. 
$\Ic$ sharply drops just below $T_{\rm c, Sr_2RuO_4}$ but increases again below a certain temperature designated as $T^*$. 
From T. Nakamura \textit{et al}. (2011).\cite{Nakamura2011PRB}
}
\label{5-2_214Pb}
\end{center}
\end{figure}

\begin{figure}
\begin{center}
\includegraphics[width=3.2in]{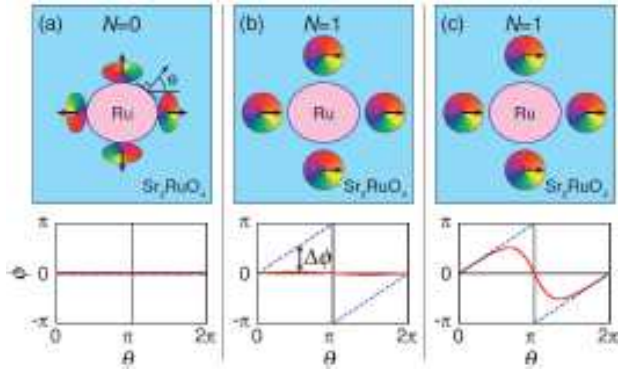}
\caption{
(Color online) 
Schematic images of the evolution of the order parameter at the \sro/Ru interface. 
In the upper panels, the gradations in the small circles depict the momentum-directional dependence of the superconducting phase $\phi$ at each spatial position; 
the arrows denote the momentum direction for which $\phi=0$. The angle $\theta$ is defined as normal to the interface (see (a)). 
The lower panels represent the superconducting phases $\phi(\theta)$ at the \sro/Ru interface under no external current. 
The solid lines represent the phase of $s$-wave superconductivity in Ru and the broken lines that of $p$-wave superconductivity in \sro. 
(a) $T_{\rm c, Sr_2RuO_4} < T < T_\varepsilon$: The $k_\parallel \pm {\rm i}\varepsilon k_\perp$ state with the winding number $N=0$ is realized (A' phase), 
matching with the $s$-wave order parameter induced in Ru. 
(b) $T \lesssim T_{\rm c, Sr_2RuO_4}$: 
Replacement by $k_x \pm {\rm i}k_y$, the bulk state of \sro, with $N=\pm1$ (B phase). 
(c) $T \ll T_{\rm c, Sr_2RuO_4}$: 
Increasing interfacial energy enlarges the phase deformation in the $s$-wave, 
strengthening the Josephson coupling. From T. Nakamura \textit{et al}. (2011).\cite{Nakamura2011PRB}
}
\label{5-2_214Pb2}
\end{center}
\end{figure}

As a novel phenomenon ascribable to real-spatial topology of chiral superconductivity, 
the critical current of a junction between \sro\ and the $s$-wave superconductor Pb has been investigated. \cite{Nakamura2011PRB} 
Pb/Ru/\sro\ proximity-junctions (Fig.~\ref{5-2_214Pb}) utilizing Ru inclusions 
in the \sro/Ru eutectic crystals as a means of achieving a good electrical contact between the two superconductors 
were fabricated. 
In such junctions, 
unusual temperature dependence of the critical current is obtained:
it increases on cooling from about 3~K but exhibits an
unusual sharp drop at the onset of bulk superconductivity of \sro\ at 1.5~K; 
at lower temperatures the critical current increases again.
This unusual behavior is ascribable to topological competition of the phase winding numbers 
around a Ru inclusion of proximity-induce $s$-wave superconductivity in Ru and of superconductivity in \sro\ as illustrated in Fig.~\ref{5-2_214Pb2}; 
thus the junction may be called a topological superconducting junction. 
The phase winding number of \sro\ changes from that of non-chiral $p$-wave superconductivity of the 3-K phase to the chiral $p$-wave superconductivity of the bulk \sro. 
We note that a similar phenomenon was reported before in Pb/\sro/Pb junctions with the critical current measured between the two Pb electrodes. \cite{Jin1999PRB}
In the recent investigation, it is demonstrated that this behavior is essentially obtained in a Pb/Ru/\sro\ junction configuration with one Pb electrode. 

Another remarkable phenomenon associated with real-space topology of superconductivity in \sro\ is 
emergence of a state related to half-fluxoid state originating from the spin and orbital degrees of freedom. 
We will describe it in the next section.

\subsection{Half-quantum vortices}\label{sec:HQV}

One of the truly remarkable phenomena expected for spin-triplet states is 
emergence of the half-quantum vortices (HQV) associated with the magnetic flux just half of the flux quantum $\Phi_0 =h/2e$. 
It was proposed in the context of superfluidity of $^3$He \cite{Volovik1976JETPL,Cross1977JLTP}, 
but has so far never been confirmed experimentally \cite{Yamashita2008PRL}. 
In general, flux quantization occurs 
as a consequence of the single-valuedness of the order parameter around a singularity: 
either a hole in a superconductor or the vortex core created in the superconductor. 
Ordinarily, this single-valuedness is satisfied 
if the orbital-phase $\varphi$ variation around a singularity, $\delta\varphi \equiv \oint \nabla\varphi\cdot\mathrm{d}\bm{s}$, is equal to $2n\pi$ with an integer $n$.
However, for a spin-triplet superconductor of the ESP type,
the single-valuedness of the order parameter may also be achieved by a combination of the variations of the spin part 
as well as the orbital part.
To demonstrate this fact, we consider an ESP spin-triplet state $\esp$ with the spin quantization axis along the $x$ axis, which can be expressed as
\begin{align}
\esp  = \sqrt{2}\varDelta e^{{\rm i}\varphi}\left(0,\,\,\cos\alpha,\,\, \sin\alpha\right)\label{eq:d-vec},
\end{align}
where the $d$-vector direction $\alpha$ corresponds to the phase of the spin part as represented in eq.~(A.2) in the Appendix.
The single-valuedness of this state is satisfied not only if $\delta\varphi=2n\pi$ and $\delta\alpha=0$ but also if $\delta\varphi=(2n-1)\pi$ and $\delta\alpha=(2m-1)\pi$, where $m$ is an integer.
The latter is an HQV, because only the orbital phase $\varphi$ couples with the vector potential and thus the magnetic flux.
The simplest example with $n=m=1$ is illustrated in Fig.~\ref{5-3_vortex}(b), in which the $d$-vector is represented by an arrow.
As we shall see in the Appendix, $\esp$ can also be expressed as 
\begin{align}
\esp = \varDelta\left( -e^{{\rm i}\tu}\uu_x + e^{{\rm i}\td}\dd_x \right),\label{eq:two-component}
\end{align}
where $\tu = \varphi-\alpha$ and $\td = \varphi+\alpha$.
Thus, an HQV can be interpreted as a vortex for which winding numbers of the components $\uu_x$ and $\dd_x$ are different.
Specifically, an HQV with the minimal vorticity, namely with $|2n-1|=|2m-1|=1$, should be coreless in the sense that one of the two components $\uu_x$ and $\dd_x$ does not have singularity at the center of the HQV, as illustrated in Fig.~\ref{5-3_vortex}(c).

\begin{figure}
\begin{center}
\includegraphics[width=3.2in]{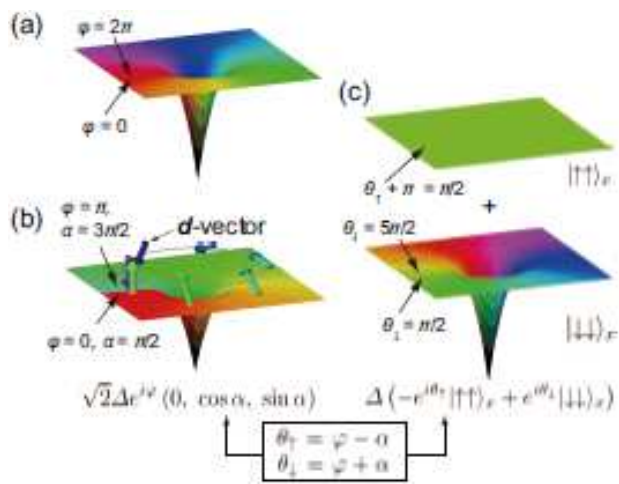}
\caption{
(Color online) Illustrations of the phases of the superconducting wave function around 
(a) a full-quantum vortex (FQV) and (b,c) a half-quantized vortex (HQV). 
As illustrated in (b), a $\pi$-rotation of the $d$-vector direction around an HQV and the $\pi$-shift of the orbital wave function recover the single-valuedness of the total wave function.
The panel (b) displays an example of an HQV in which the phases of the order parameter change from ($\varphi$, $\alpha$)=(0, $\pi/2$) to ($\pi$, $3\pi/2$). In this example, $\dv$ rotates from $\dv\parallel \zh$ to $\dv \parallel -\zh$.
An HQV can be also interpreted as a vortex for which winding numbers of the components $\uu_x$ and $\dd_x$ are different.
An example equivalent to the case in (b) is shown in (c).
Here, $\tu=\varphi-\alpha$ is constant whereas $\td=\varphi+\alpha$ changes from $\pi/2$ to $5\pi/2$.
Note that the actual phase for $\uu_x$ is $\tu+\pi$ due to the minus sign in the first term in \eq{eq:two-component}.
}
\label{5-3_vortex}
\end{center}
\end{figure}

The HQV involves the rotation of the $d$-vector and the associated spin currents expressed as $j_s \propto \nabla\alpha$, 
in which $\alpha$ is the directional angle of the $d$-vector. Thus the total energy associated with an HQV is
\begin{equation}
E_\mathrm{HQV}=C[-\rho_\mathrm{charge}(\nabla \varphi)^2+\rho_\mathrm{spin}(\nabla\alpha)^2]
\end{equation}
with a coefficient $C$.
The charge current associated with an HQV is screened in a distance characterized by the London penetration depth, 
whereas there is no such screening effect expected for the spin current. 
To limit the increase of the energy due to spin current, 
it is anticipated that either a mesoscopic sample size or a closely-separated pair of HQVs is necessary \cite{Chung1997PRL}. 
In the former, when the size of a sample is comparable to the penetration depth, 
HQVs are similar in energy to full-quantum vortices (FQVs). 

\begin{figure}
\begin{center}
\includegraphics[width=3.2in]{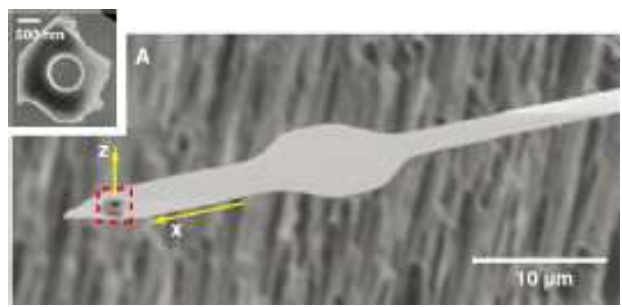}
\caption{
(Color online) Silicon cantilever used in an ultra-high-sensitivity torque magnetometer 
with a micron-size crystal ring of \sro\ glued. 
The Sr$_2$RuO$_4$ ring is also shown in the inset.
From Jang \textit{et al}. (2011).\cite{Jang2011Science}
}
\label{5-3_cantilever}
\end{center}
\end{figure}

In search of an HQV, measurements of magnetic torque of micron-size rings of \sro\ have recently been performed. 
As shown in Fig.~\ref{5-3_cantilever}, a micrometer-size crystal with a hole drilled by an FIB technique 
was placed on a silicon cantilever of laser-detected ultra-sensitive torque magnetometer \cite{Jang2011Science}. 
The purpose of the central hole is to avoid the complications associated with vortex core production. 
In such a small sample, the charge supercurrent is non-zero everywhere and it is a fluxoid, 
given by the sum of flux and kinematic term, rather than a flux, that is quantized to an integer multiple of flux quantum: 
\begin{equation}
\Phi+\frac{m}{e} \oint \mib{v}_s \cdot \mathrm{d} \mib{s} = n \Phi_0
\end{equation}
With increasing the magnetic field penetrating the hole, 
the magnetic moment of the ring is found to decrease uniformly with a negative slope corresponding to the Meissner screening. 
In addition, a positive jump is observed periodically, corresponding to the full fluxoid quantization. 
The interval of these jumps is consistent with the fluxoid quantization given by eq.~(10).

\begin{figure}
\begin{center}
\includegraphics[width=3.2in]{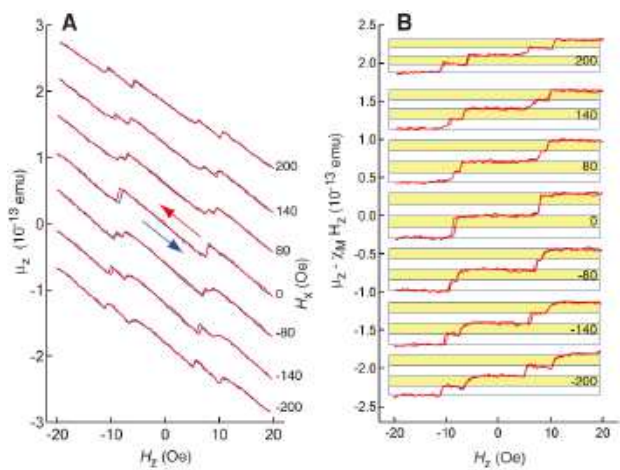}
\caption{
(Color online) (A) Magnetization of a micron-size ring of a \sro\ crystal under magnetic fields $H_z$, 
showing a slope due to Meissner current as well as exhibiting jumps associated with fluxoid quantization. 
With additional fields perpendicular to the ring axis, the fluxoid jumps split into two. 
(B) Magnetization of the ring sample after the Meissner shielding contribution has been subtracted. 
The magnitude of magnetization steps of split jumps is nearly half of that corresponding to a full fluxoid jump. 
The plateau thus is consistent with a formation of a half-fluxoid state, closely related to the HQV state. 
From Jang \textit{et al}. (2010).\cite{Jang2011Science}
}
\label{5-3_HIstate}
\end{center}
\end{figure}

Unusual jumps between the FQV jumps are observed when additional field is applied 
perpendicular to the axis of a hole as in Fig.~\ref{5-3_HIstate}. 
The range of longitudinal field for the stability of the half-fluxoid regions increases 
nearly proportionally to the perpendicular field. 
Experimentally, the size of each magnetization jump as well as the magnitude of the longitudinal field 
is precisely consistent with the half fluxoid expectation. 
Other possible reasons of the fractional magnetization jumps were examined. 
For example, penetration of quantized vortices 
through the side walls may result in the magnetization jumps of varied magnitude. 
To test this possibility, the perpendicular field is applied to different in-plane direction on the same crystal, 
but the character of the magnetization jumps remains unchanged: it always occurs as half-fluxoid jumps. 
The measurements were repeated in several crystals: 
the HQV feature was not detected in a crystal with the size greater than a few microns, 
or in NbSe$_2$ crystals with a comparable anisotropy as \sro.

The origin of the necessity of the perpendicular fields has not been fully understood. 
A semi-quantitatively feasible mechanism is due to ``kinematic polarization'' 
proposed by Vakaryuk and Leggett \cite{Vakaryuk2009PRL}. 
In an HQV state, the orbital winding numbers of the spin-up and spin-down species differ by unity. 
The associated difference in the kinetic energy due to unbalanced charge current 
results in the unequal population of each spin species. 
Application of the field parallel to the $ab$ plane would then lower 
the spin-polarization energy of an HQV state compared with that of an FQV state. 
In order to establish the HQV state, other direct measurements of the flux quantization are much desired. 
In any case, the observations of half-height flux jumps by Jang \textit{et al}. certainly give 
additional strong confirmation of the spin-triplet ESP state realized in \sro.

\section{Establishing spin-triplet superconductivity}

We have reviewed the current status of our knowledge of unconventional superconductivity in \sro. 
From the material perspective, the investigations are greatly helped 
by the availability of 
large size, high-quality single crystals:
they are chemically-stable, stoichiometric, and having the quasi-particle mean free path 
more than an order of magnitude greater than the superconducting coherence length.
The superconducting eutectic systems containing \sro\ add opportunities to study novel superconducting properties 
including exotic proximity effects. 
From the viewpoint of the electronic structure, 
strongly-correlated multiband Fermi surfaces are relatively simple thanks to the structural simplicity, 
and the normal state properties are quantitatively well characterized as a Fermi liquid. 
This allows development of microscopic mechanism theories of unconventional superconductivity 
based on the realistic multiband Fermi surfaces.

From the extensive NMR studies, the realization of the spin-triplet pairing seems quite certain. 
Nevertheless, obtaining the complete picture of superconductivity of \sro\ still remains as a challenge. 
As a promising future direction, we have emphasized the necessity of including spin-orbit interactions 
in conjunction with three dimensional aspect of the electronic states.
Inclusion of these factors may lead to improved interpretations of the mechanism of superconductivity 
as well as crucial understanding of some of the unresolved issues.
To advance further, it is necessary to revisit the superconducting double transitions near $\Hcc$ 
as well as the collective-mode excitations observed in the NMR relaxation under zero-magnetic field. 
Relatively unexplored areas also include the properties of vortex lattice 
especially with the field parallel to the basal plane. 

A new perspective as a ``topological superconductor'' should also be important. 
Generation of half-quantum vortices, observation of 
phenomena associated with chiral edge states, 
fabrication of superconducting junctions exhibiting novel behavior 
such as odd-frequency pairing states are some of the promising future directions. 
Studies using micron-size crystals are proven to be essential to uncover some of these phenomena. 
At the same time, in order to firmly establish the chiral $p$-wave state with the two-component order parameter, 
superconducting properties under symmetry-breaking fields, 
for example, in-plane uniaxial pressure, may provide useful information.

The extensive efforts to elucidate the superconductivity of \sro\ 
in all details demonstrate the possible approaches that 
can be taken to unveil novel behaviors of an unconventional superconductor. 
In this sense, we hope that this review also serves useful for the studies of other unconventional superconductors. 

\acknowledgement
The authors would like to thank 
M.~Sigrist, K.~Yamada,  A.~P.~Mackenzie, K.~Deguchi, Y.~Yanase, S.~Kashiwaya, Y.~Tanaka, Y.~Asano, 
H.~Ikeda, H.~Yaguchi, A.~Kapitulnik, Z.~X.~Shen, K.~Molar, C.~Kallin, 
Y.~Liu, K.~Miyake, Y.~Hasegawa, K.~Machida, T.~Ishiguro, R.~Budakian, D.~F.~Agterberg, A.~J.~Leggett, T.~M.~Rice,
A.~Vecchione, R.~Fittipaldi, Taketomo Nakamura, T.~Terashima, I.~Vekhter, V.~P.~Mineev, J.~A.~Sauls, and J.~Goryo
for valuable communications. 
This work was supported by the ``Topological Quantum Phenomena'' (22103002) and ``Heavy Electrons'' (20102006, 23102705) Grant-in-Aid for Scientific Research on Innovative Areas, 
a Grant-in-Aid for the Global COE program ``The Next Generation of Physics, Spun from Universality and Emergence'', and 
Research Activity Start-up (22840013) from the Ministry of Education, Culture, Sports, Science and Technology (MEXT) of Japan and 
the Japan Society for the Promotion of Science.

\appendix
\section{Two equivalent interpretations of the half-integer quantum vortex}

As briefly mentioned in \S~\ref{sec:HQV}, there are two equivalent interpretations of the half-integer quantum vortex (HQV): 
One is that $(2n-1)\pi$ phase shift in the orbital part of the order parameter and additional $(2m-1)\pi$ phase shift in the spin part around an HQV. 
The other interpretation is a vortex for which winding numbers of the two ESP components are different.
In this Appendix, we explain the equivalence of these two interpretations.

First, we write again the definitions of the bases of the $d$-vector:
\begin{align*}
\xh&\equiv \bra{S_x = 0} = \frac{1}{\sqrt{2}}\left(-\uu_z + \dd_z\right), \\
\yh&\equiv \bra{S_y = 0} = \frac{{\rm i}}{\sqrt{2}}\left(\uu_z + \dd_z\right),  \\
\zh&\equiv \bra{S_z = 0} = \frac{1}{\sqrt{2}}\left(\ud_z + \du_z\right),
\end{align*}
where the subscript $z$ in the state vectors indicates that the spin quantization axis is the $z$ axis.
If one takes $x$ as the spin quantization axis, an alternative equivalent set is obtained as
\begin{align*}
\xh&\equiv \bra{S_x = 0} = \frac{1}{\sqrt{2}}\left(\ud_x + \du_x\right),\\
\yh&\equiv \bra{S_y = 0} = \frac{1}{\sqrt{2}}\left(-\uu_x + \dd_x\right),\\
\zh&\equiv \bra{S_z = 0} = \frac{{\rm i}}{\sqrt{2}}\left(\uu_x + \dd_x\right).
\end{align*}

Here, we start with a spin-triplet equal-spin pairing (ESP) state $\esp$ with the quantization axis along $x$:
\begin{align}
\esp=\varDelta\left(-e^{{\rm i}\tu}\uu_x + e^{{\rm i}\td}\dd_x\right).
\end{align}
We note that the minus sign in the first term is just for a convenience that the state with the zero phases $\tu=\td=0$ corresponds to the $\yh$ state.
Then, we can convert them to another set of phases: $\alpha \equiv (\td-\tu)/2$ and $\varphi \equiv (\td+\tu)/2$. 
Using $\alpha$ and $\varphi$, $\esp$ can be expressed as 
\begin{align}
\esp = \varDelta e^{{\rm i}\phi}\left(-e^{-{\rm i}\alpha}\uu_x + e^{{\rm i}\alpha}\dd_x \right)\label{eq:app:esp-1}.
\end{align}
This equation can be further modified using the relations $\uu_x = -(\yh+{\rm i}\zh)/\sqrt{2}$ and $\dd_x = (\yh-{\rm i}\zh)/\sqrt{2}$ as
\begin{align}
\esp &= \frac{\varDelta e^{{\rm i}\varphi}}{\sqrt{2}}\left[ (e^{{\rm i}\alpha}+e^{-{\rm i}\alpha})\yh + (e^{{\rm i}\alpha}-e^{-{\rm i}\alpha})(-{\rm i}\zh)\right] \nonumber\\
& = \sqrt{2}\varDelta e^{{\rm i}\varphi}\left(0,\,\,\cos\alpha,\,\, \sin\alpha\right)\label{eq:app:d-vec}.
\end{align}
Thus, this relation indicates that $\varphi$, the average of the phases of the up- and down-spin components, corresponds to the phase of the orbital part of the order parameter, 
and that $\alpha$, the half-difference between $\tu$ and $\td$, corresponds to the direction of the $d$-vector in the plane perpendicular to the quantization axis.

When an ordinary full-quantum vortex (FQV) exists, the orbital phase $\varphi$ must shift by $\pm2n\pi$ after a circulation around the vortex in order to satisfy the single-valuedness of the wavefunction. 
According to \eq{eq:app:d-vec}, the single-valuedness is also satisfied if the phases $\alpha$ and $\varphi$ both additionally shift by $\pm\pi$. 
This is the HQV, because only the shift in the orbital phase $\varphi$ couples to the vector potential resulting in the magnetic flux of $\varPhi_0/2 = h/4e$.
As one can see, the $\pi$ phase shift in $\alpha$ around an HQV corresponds to a rotation of the $d$-vector by $180\deg$ around the quantization axis. 
These facts are illustrated in Fig.~\ref{5-3_vortex}(b).
In an alternative viewpoint, the phase for the up-spin component $\uu_x$ is $\tu=\varphi-\alpha$ and the phase for $\dd_x$ is $\td=\varphi+\alpha$ as shown in \eq{eq:app:esp-1}. 
Thus, the winding numbers of $\tu$ and $\td$ are different around an HQV.
A remarkable example is the case that the shifts in $\alpha$ and $\varphi$ around an HQV are both $\pm\pi$.
In such cases, only one of $\uu_x$ and $\dd_x$ exhibits a phase shift of $\pm2\pi$, whereas the other exhibits zero shift. 
Thus, one of the two components $\uu_x$ and $\dd_x$ has zero vorticity around an HQV as schematically explained in Fig.~\ref{5-3_vortex}(c). 
Such an HQV is ``coreless'' in a sense that one of the two components has no singularity at the center of the HQV, if it is realized in a simply connected bulk sample.

\end{document}